\documentclass{aa} 

\usepackage{graphicx}
\usepackage{lscape}
\usepackage{longtable}
\usepackage{rotating}
\usepackage[comma,authoryear]{natbib}
\usepackage{txfonts}
\newcommand{\Mjup}{M$_{\rm Jup}$\,}
\newcommand{\Mearth}{M$_{\rm Earth}$}
\newcommand{\mjup}{M$_{\rm Jup}$\,}

\newcommand{\vsini}{$v\sin${\it i}\,}

\newcommand{\harps}{H{\small ARPS}}

\newcommand{\bp}{$\beta$\,Pictoris\,}
\newcommand{\bpic}{$\beta$\,Pictoris\,}
\newcommand{\Ld}{L$_{\rm disk}$}
\newcommand{\Ls}{L$_{\rm star}$}

\begin{document}

   \title{Planets around stars in young nearby associations
     \thanks{Based on observations made with the ESO3.6m/Harps spectrograph at La Silla.083.C-0794(ABCD); 084.C-1039(A); 084.C-1024(A)}
     \thanks{Tables of radial velocities are only available in electronic form at the CDS via anonymous ftp to cdsarc.u-strasbg.fr (130.79.128.5) or via http://cdsweb.u-strabg.fr/cgi-bin/qcat?J/A+A/}}

   \subtitle{Radial Velocity searches: a feasibility study, and first results }

   \author{  
     A.-M. Lagrange \inst{1} \and
N. Meunier \inst{1} \and
  G. Chauvin \inst{1} \and	
   M. Sterzik \inst{2} \and	
F. Galland \inst{1} \and	
G. Lo Curto	\inst{3} \and
J. Rameau \inst{1} \and
D. Sosnowska \inst{4} 
  }

   \institute{
     Institut de Plan\'etologie et d'Astrophysique de Grenoble, UMR5274 CNRS, Universit\'e Joseph Fourier, BP 53, 38041 Grenoble Cedex 9, France\\
     \email{anne-marie.lagrange@obs.ujf-grenoble.fr}
\and
European Southern Observatory, Casilla 19001, Santiago 19, Chile
\and
European Southern Observatory, Karl-Schwarzschild-Str. 2, D-85748 Garching bei M¨unchen, Germany
\and
Observatoire de Gen\`eve, Universit\'e de Gen\`eve, 51 ch. des Maillettes, CH-1290 Versoix, Switzerland
}

   \date{Received date: 21 Nov. 2012 / Accepted date : 21 May 2013}

   
   \abstract
   { Stars in young nearby associations are the only targets allowing giant planet searches at all separations in the near future, by coupling indirect techniques such as radial velocity and deep imaging. These stars are first priorities targets for the forthcoming planet imagers on 8 to 10 meter class telescopes. Young stars rotate more rapidly and are more active than their older counterparts. Both effects can limit the capability to detect planets using RV.
}
   { We wish to explore the planet detection capabilities of a representative sample of stars in close and young associations with radial velocity data  and explore the complementarity between this technique and direct imaging.
   }
   { We observed 26 such targets with spectral types from A to K and ages from 8 to 300 Myr with \harps. We compute the detection limits  with two methods, in particular, a method we recently developped, that takes into account the frequency distribution of the RV variations. We also attempt to improve the detection limits in a few cases by  correcting for the stellar activity.
   }
   { Our A-type stars RV show high frequency variations due to pulsations, while our F-K stars clearly show  activity with more or less complex patterns. For F-K stars, the RV jitter and \vsini rapidly decrease  with star age. The data allow us to search for planets with periods typically ranging from 1 day to 100 days, and up to more than 500 days in a few cases. Within the present detection limits, no planet was found in our sample. For the bulk of our F-K stars, the detection limits fall down to sub-Jupiter masses. We show that these limits can be significantly improved by correcting  even partially for stellar activity, down to a few Neptune masses for the least active stars. The detection limits on A-type stars can be significantly improved, down to a few Jupiter mass planets, provided an appropriate observing strategy. We finally show the tremendous potential of coupling RV and AO deep imaging results. 
	}
{The RV technique allows the detection of planets lighter than Jupiter, and down to a few Neptune masses around young stars aged typically 30 Myr or more. Detection limits increase at younger ages, but (sub-)Jupiter mass planets are still detectable. In the forthcoming years, the use of complementary techniques will allow a full exploration of the Jupiter mass planets content of many of these stars.
}

   \maketitle
%

\section{Introduction}


Thanks to hundreds of  discoveries (http://exoplanet.eu/) for more than 15 years, mainly coming from radial velocity (RV) and transit surveys, our knowledge of exoplanets has dramatically improved. We know that exoplanets are frequent around solar-type, main sequence (MS) stars: more than 50$\%$ have planets \cite[all masses,][]{mayor11} and about 15$\%$ have planets with masses larger than 50~\Mearth. An unexpected diversity of planet properties such as separations, eccentricities and orbital motions (for example retrograde orbits discovered thanks to the Rossiter-McLaughlin effect) was revealed for short or intermediate period planets. This suggests that dynamics, due for example to planet-planet interactions or early disk-planet interactions, plays an important role in the building of planetary systems. 

Giant planets play an important role as they carry most of the planetary system mass. They therefore strongly impact the dynamics and fate of lighter planets and the final architecture of the planetary systems. Giant planets also impact the detectability of lighter planets as far as indirect methods are concerned. Even though giant planets represent the majority of the planets detected so far, we are far from having a clear picture of their occurence,  variety and  properties. RV and transit explorations are indeed still limited to planets orbiting within 5 AU from their parent stars, while current deep (adaptive-optics, herefater AO) imaging, meant to detect more distant planets, are not sensitive enough for MS stars, except around a  few young early-type stars which have already reached the MS. Stars in young nearby associations \cite[see for example][]{torres08,zuckerman04} are therefore the best targets for a complete giant planet exploration, combining the data obtained with forthcoming planet imagers such as SPHERE at the VLT \cite[][]{beuzit08} or GPI at GEMINI \cite[][]{macintosh08} and high precision spectrographs. In addition, these data could be combined in the future with astrometric measurements when available. 

 Compared to mature, MS stars, limited effort has been devoted so far to the search for planets around young stars with RV  techniques. To our knowledge, the only significant surveys were performed by \cite{paulson04} on 94 stars members of the Hyades (aged about 600 Myr) and on 61 stars members of  various moving groups (MGs) or stellar associations aged between 12 and 300 Myr \cite[][]{paulson06}. In both cases, high spectral resolution data, in the range 40000-70000 were used. The  internal errors were  3-5 and $\simeq$ 10 m/s. The Hyades stars were found to be weakly active, with stellar jitters in the range 10-20 m/s (average value = 16 m/s). No planet was detected with masses down to 1-2 MJup, and down to the sub-Jupiter mass regime in a few cases and with periods typically of 1-2 years or less. In the second case, no planet was detected down to 1-2 MJup, for periods up to 6 days for the oldest stars while the detection limits fell into the brown dwarf regime for the youngest targets (members of the $\beta$ Pic group). 

Stellar activity can mimic planet signatures \cite[][]{desort07}, and great care must be taken when analysing RV data. In the past, two detections were announced  around TW Hya \cite[][]{setiawan07}, and BD +20 1790 \cite[][]{hernan10}, but were strongly debated  \cite[][]{figueira10}.  \cite{bailey12} used NIRSPEC at Keck to observe a set of 20 young stars with high precision spectroscopy in the near-IR, where line shifts due to spot-induced activity are expected to be smaller than at optical wavelengths. Detection limits of 8~\Mjup (respectively 17 \Mjup) were obtained for periods of 3 days (respectively 30 days) periods. \cite{Crockett12} used CSHELL for similar purposes on 9 T Tauri stars and showed that for these stars, the jitter in the near IR was reduced by a factor 2-3 compared to the jitter measured in the optical range. No planet was firmly detected.  Finally, we detected unambiguously a  2.8 \mjup giant planet orbiting with a 320 days period, a 150 Myr, weakly active, star \cite[][]{borgniet13}. 

We investigate here the potential of high precision optical spectroscopy to search for giant planets around 26 stars which are members of close-by young associations, as such stars are potential targets of the SPHERE NIRSUR survey aimed at imaging giant planets. Section 2 describes our targets and observing log. Section 3 presents the data analysis. The results are shown and discussed in Sect. 4. In Section 5, we show how such results can be coupled to deep imaging detection limits to perform a complete exploration of giant planets in the environements of these young and close stars.

\section{Targets and observations}

  \subsection{Target list}
The bulk of our sample is made of 22 bright ($V\le9$), nearby ($d \le 80$~pc), young ($\leq$ 200 Myr) stars identified in young MGs and from systematic spectroscopic surveys. Four other stars with ages greater than 200 Myr were observed: HD90905, HD109536, HD105690 and HD188828. The 26 stars are listed in Table~\ref{targets}. Our sample includes 8 A-type stars, 4 F-type stars, 10 G-type stars and 4 K-type stars. Several of them are known to have debris disks \cite[][]{rodriguez12}. In several cases,  peculiar disks properties (shapes, offsets) suggest that inner planets could be present.

We provide now some details about a few peculiar targets. Otherwise indicated, relative disk luminosities \Ld/\Ls$\;$ are from \cite{moor06} or \cite{rhee07}, and binary indications from \cite{rodriguez12}:
\begin{itemize}
\item[-] HD39060 (\bp) is surrounded by a seen edge-on disk (\Ld/\Ls = 2.10$^{-3}$), as well as a planet detected in direct imaging \cite[][]{lagrange09a,lagrange10}. Detection limits in deep imaging fall in the giant planet regime. A study of RV data allowed \cite{lagrange12} to set an upper limit to the dynamical mass of \bp b. 
\item[-]HD61005 (The Moth) is surrounded by a thin and narrow disk which is resolved in near-IR \cite[][and references therein]{hines07,buenzli10}. The disk is seen as a ring inclined by 84 degrees with respect to pole-on and is offset from the star center. In addition, streamers are observed at the edge of the ring. Detection limits in deep imaging fall in the giant planet regime. 
\item[-] HD172555 belongs to the \bp MG \cite[][]{zuckerman01}. It has a spectral type similar to that of \bpic but an IR excess smaller than that of \bp (\Ld/\Ls = 8.10$^{-4}$). It is also part of a wide binary system. 
\item[-] HD174429 (PZ Tel) has an IR excess (\Ld/\Ls = 9.10$^{-5}$). A brown dwarf companion with a 20-40 \Mjup mass is present with a projected separation of 0.3" representing 15 AU \cite[][]{biller10}. Given the companion $\Delta$J (5.6 mag), and its expected V-J (5.5-6.5 mag) given its mass and age, and the V-K (1.2 mag) of the star itself, the contrast at optical wavelengths between PZ Tel and its companion is about 9-10 mag. The companion contribution to the visible flux within the Harps fiber (1") is therefore negligeable. 
\item[-] HD181327 is a member of the \bp MG, with a large IR excess \cite[\Ld/\Ls = 2.10$^{-3}$][]{lebreton12}. A debris disk has been resolved in near IR by \cite{schneider06} as a ring located at 90 AU and inclined by about 30 degrees from face-on. 
\item[-] HD216956 (Fomalhaut) is also surrounded by a debris disk (\Ld/\Ls = 8.10$^{-5}$) seen as a narrow ring at 115 AU \cite[][]{kalas05}, inclined by 24 degrees from edge-on and offset from the star. It may also be surrounded by a planet orbiting just inside the ring \cite[][]{kalas08}, and shaping its sharp inner edge \cite[][]{chiang09}, but this planet is still debated \cite[][]{janson12}. Recently \cite{lebreton13b} showed that warm dust was also present within AUs from the star.
\item[-] HD218396 (HR8799) is surrounded by a system of 4 planets detected in imaging with projected separations between 15 and 70 AU \cite[][]{marois08,marois10}. A thin debris disk (\Ld/\Ls = 2.10$^{-4}$) has also been resolved at sub-mm wavelengths \cite[][]{patience11} and has an inclination of 20-50 degrees from face-on \cite[][]{kalas10}.
\end{itemize}

  \subsection{Observations}

We have obtained more than 2000 high resolution ($R \simeq 115000$) and high signal to noise ($\mathrm{S/N}$) spectra of our targets  with the fiber-fed spectrograph Harps \cite[][]{mayor03} at the La Silla 3.6m telescope. In the case of $\beta$ Pictoris,  we used in addition to our data, a few data available in the archive and obtained with the same set-up.  The spectra cover a wavelength range between 3800 and 6900\,$\AA$. During our runs, we usually recorded two consecutive spectra for each telescope pointing. In the following, one pointing will be refered to as one "epoch". As our main goal was to look for short period planets (100 days or less), we recorded several spectra of each target during each run whenever possible. In addition, we also recorded whenever possible spectra at two or three different epochs during one night, to identify possible high-frequency RV variations. 
 For the brightest stars, we did not consider  spectra with saturation that could affect the RV measurements. We report in Table~\ref{obs} information about each star observation. The time-span varies between a few hours for 3 of our targets  and 1912 days. The stars with a long time-span are early-type stars which happened to be part of a previous survey on A-F stars described in \cite{lagrange09b}. The exposure times were computed to get a $\mathrm{S/N}$ larger than 200 at 550 nm in most cases. This corresponds to exposure times between less than 1 minute for the brightest targets and 15 minutes for the faintest ones.

Finally, for some of the earliest types, generally pulsating stars, we performed continuous observations over typically 1-2 hours to check that short-term pulsations are indeed present and to estimate their amplitude.

 \section{Data analysis }

\subsection{RV measurements}

The RV were computed using SAFIR, a tool dedicated to measure accurate RVs on fast rotators, described in \cite{galland05}, and based on the Fourier interspectrum method developed by \cite{chelli00}. SAFIR has been extensively used, in particular in RV surveys of rapidly rotating stars \cite[][]{lagrange09b}. SAFIR input data are the extracted 2D spectra provided by the Data Reduction Software (DRS) pipeline, and the related wavelength calibration coefficients. SAFIR takes care of the blaze correction as well as bad-pixel correction \cite[][]{galland05}. The errors associated to the individual RV measurements range between typically 1.5 m/s for the slowly rotating stars to typically 30-40 m/s for the fast rotating stars. The values obtained for slowly rotating stars are similar to those obtained by the HARPS DRS; these values show that the instrument internal errors are very low. Lomb-Scargle periodograms of the RV variations were also computed for each target. 

 


\subsection{Line profile variations and activity indicators} 

\begin{figure*}[htp!]
    \centering
\includegraphics[angle=0,width=1.\hsize]{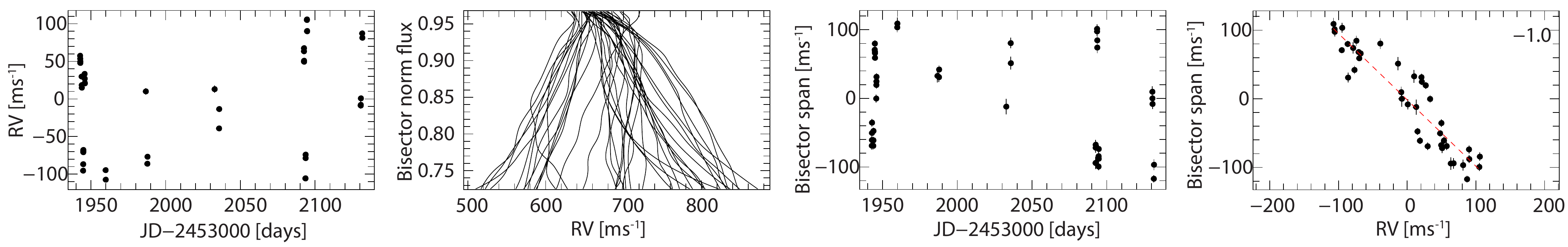}
\includegraphics[angle=0,width=1.\hsize]{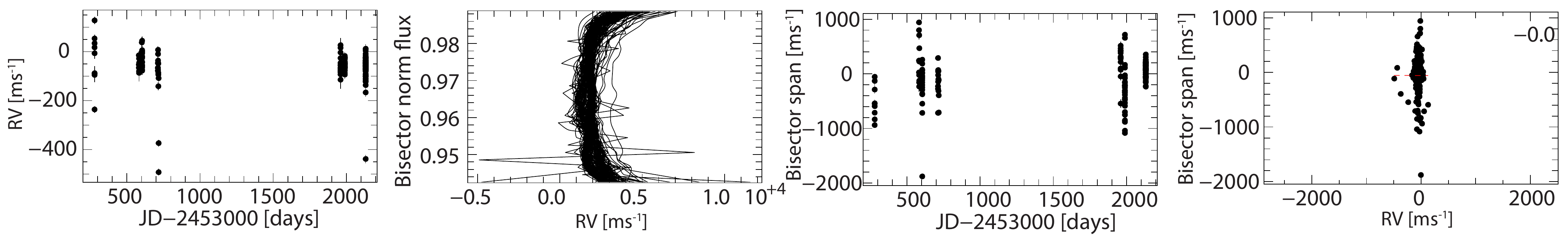}
    \caption{{\it Upper panels}: RV versus time, bisectors, BVS versus time and BVS versus RV for the active G-type star HD183414. In the last panel, we indicate the slope of the (RV, BVS) correlation. {\it Lower panels}: same for the pulsating A-type star HD216956. }
    \label{bsv}
  \end{figure*}

For stars with \vsini below typically 100 km/s and spectral types later than A3V, it is possible to compute the bisector velocity spans (BVS) using the cross-correlation function, as shown in \cite{lagrange09b}. Typical examples are provided in Fig.~\ref{bsv} for an active star (upper panels) and a pulsating one (lower panels). The relation between BVS and RV is useful to further characterize the temporal RV variations whenever present. A correlation between the RV and BVS variations indicates that the RV are due to stellar activity \cite[][]{desort07} while pulsation-dominated variations induce instead uncorrelated (RV, BVS) variations. We also computed the \vsini using the cross-correlation function for the low to moderately rotating stars and the obtained values are given in Table~\ref{obs}. 

For active stars, we also computed the S-index that measures the amount of chromospheric activity, and derived the Ca index R'$_{HK}$ according to \cite{duncan91}\footnote{Note that we did not apply any correction factor to the obtained R'$_{HK}$ index, as it is sometimes done to calibrate the values with those of Mount Wilson, because 1) we are mainly interested here in relative variations, and 2) such corrections are possible only when a lot of values (for several targets, with stable levels of activity), taken both at Mount Wilson and in the other observing site are available. Here, we are possibly dealing with a strong variability, so a correction could be applied only if we had simulatenous measurements, for the same targets. Our log(R'$_{HK}$) are similar within 0.3 dex with those already published ones however.}. In the case of the Sun, there is a correlation between the long-term Ca emission variability and the RV variability \cite[see][and references therein]{meunier13}. For solar-type MS stars, a correlation between the long-term photometric variability and the Ca emission is also observed \cite[see in particular][]{lockwood07}: the brightness increases when the Ca emission increases. This is consistent with the fact that the long-term photometric variations are, as for the Sun, dominated by plages. Correlations between RV and Ca emission have been searched for \cite[][]{santos10,isaacson10,boisse09,dumusque11}, but only rarely reported \cite[][and references therein]{dumusque11}. However, instrumental artefacts such as  diffused light could alter the Ca emission measurements and/or the RV measurements (the long term stability must be ensured to a few m/s) and may therefore prevent finding such a correlation. For  young solar-type stars, the situation is even less clear, as much less data are available. The long-term photometric variations are anti-correlated with chromospheric activity variations, i.e. the brightness decreases when the Ca emission increases, in contrast to older Sun-like stars \cite[][]{lockwood07}. This is interpreted as the photometric variations being spot-dominated.  

For stars  with a high enough \vsini (i.e. larger than 15-20 km/s) or with very strong emissions, the 1~$\AA$ wavelength range used to measure the Ca emission index is too narrow compared to the emission itself. Even though sophisticated modeling has been proposed to estimate the Ca index in such cases \cite[][]{schroeder09}, we did not make any attempt to do so, as the main purpose of the present paper is not the study of stellar activity. In such cases no value is given in Table~\ref{obs}.

\subsection{Detection limits measurements}

\begin{figure*}[htp!]
    \centering
\includegraphics[angle=0,width=1.\hsize]{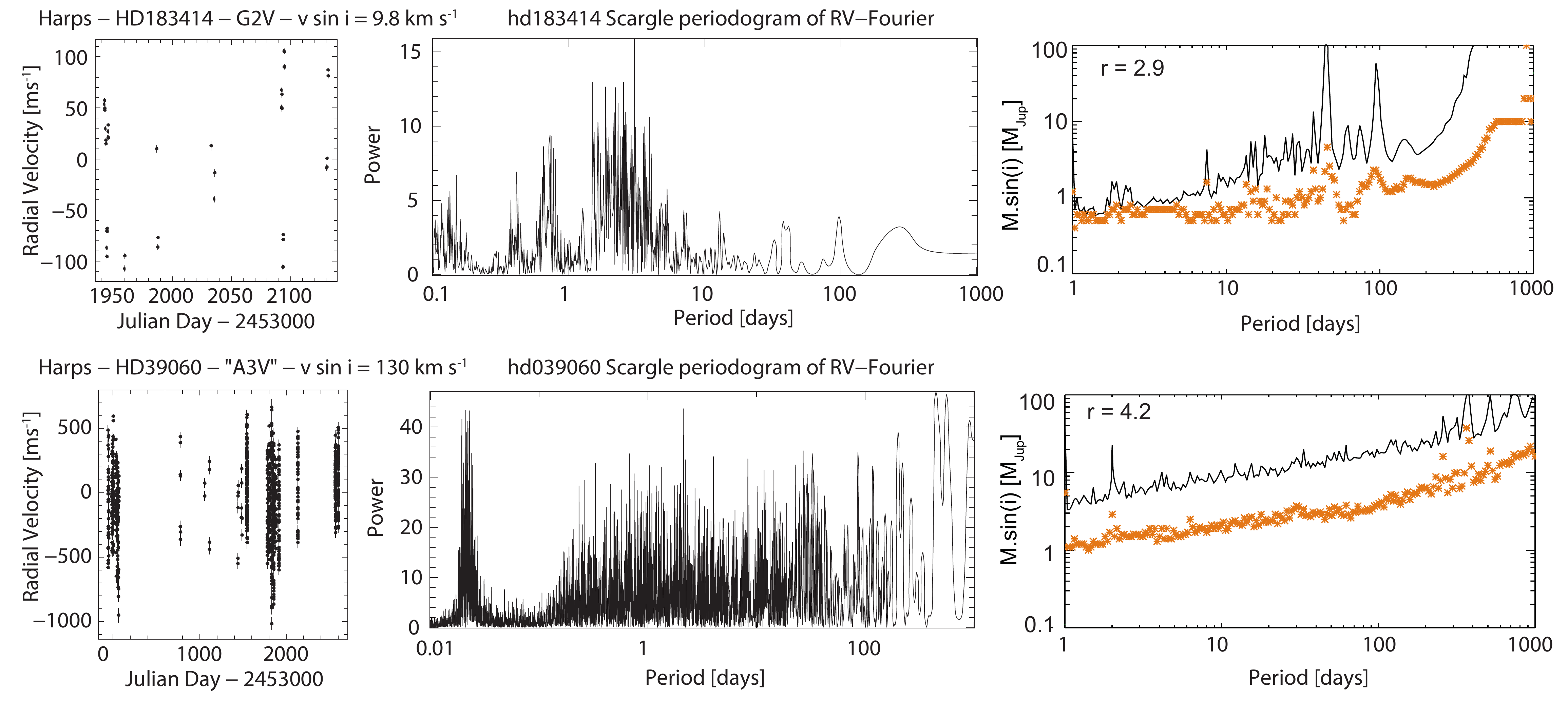}
    \caption{{\it Upper panel}:  RV versus time, periodogram of RV and detection limits [{\it from left toright}) for the G2V star HD183414, covering a time span of 189 days. The detection limits have been computed using the rms-based method (black line), LPA (orange stars). {\it Lower panels}: same for the A5V star HD39060 covering a time span of 2600 days.}
    \label{examples_limdet}
  \end{figure*}

As extensively discussed in a previous dedicated paper \cite[][hereafter MLB12]{meunier12}, there are different ways to estimate detection limits associated to RV time series. We will use two methods in the present paper, namely the  "rms-based method" and the LPA method. Noticeably, we will not use the bootstrap method which, we showed, provides detection limits not as good as the LPA method and in the case of pulsating stars, may provide unrealistic results. We refer to MBL12 for a detailed description of these methods and on their respective merits. We briefly summarize the principles of the "rms-based" and LPA methods. Examples of obtained detection limits are provided in Fig.~\ref{examples_limdet}.

The fastest method, the "rms-based method" is based on the measured standard deviation (hereatfer rms) of the RV serie and was first described in \cite{lagrange09b}. In the rms-based method, we compute the RV produced by a planet with a given mass and period for the actual temporal sampling and then the rms of these simulated RV. This is done for 1000 orbital phases. The distribution of this rms is Gaussian. If the rms of the observed RV is lower than the average rms of the 1000 simulated RV, the planet is detectable. The level of confidence (detection probability) is obtained by comparing the standard deviation of the simulated distribution and the difference between the observed rms and the average value of the simulated distribution.  We showed in \cite{lagrange09b} that provided a very good temporal sampling, the rms-based detection limit in mass for a given period is similar to the mass of a planet that would produce an RV amplitude equal to 3$\times$rms(RV). In the following,  the later limit will be referred to as the rms-based achievable limit.

The LPA (local power analysis) method provides more accurate and lower detection limits when enough data are available (see below). Briefly, for a given period P, we compute the maximum of the periodogram of the observed RV in the range 0.75P-1.25P\footnote{We checked that considering different period ranges, such as 0.7P-1.3P, 0.8P-1.2P, 0.9P-1.1P does not significantly change the detection limits curves.}: this provides a threshold for that period. Then, for a given planet (mass, period), we compare the maximum amplitude of the power spectrum of the RV induced by this planet on the same temporal sampling with that threshold, for 100 trials of the phase of the planet. If for all these trials, all "planet amplitudes" are above the threshold then the planet mass is above the detection limit. We iterate on the planet mass  until we reach a mass for which this condition is  no longer met, indicating a mass below the detection limit. The smallest step we use is in general 0.1 \Mjup, but we chose a smaller step in a few cases to avoid a discretization of the detection limits.
 With our present set of data, we can use the LPA method for ten stars, for which more than 40 data points were available.



As expected, the LPA detection limits are much lower than the rms-based limits. To quantify the  improvement, we compute the ratio between the rms-based and the LPA detection limits for each of the 200 periods considered, and then the median of the 200 ratio. The median ratio is typically a factor of 2 to 3, but may sometimes reach higher values (see Section 4). The improvement is due to the fact that the LPA method takes into account the variability of the power amplitude frequency, either due to an actual signal or to the temporal sampling, while the rms-based does not.
The advantage of the rms-based is that the computation is fast, and it is  less sensitive to the number of data available. It provides an order of scale but pessimistic estimation of the achievable detection limits \cite[][]{lagrange09b}. 

\subsection{Search for circumstellar Ca in rapid rotators}

Finally, as a by-product of these observations, we also used the spectra to search for narrow CaII absorption lines at the bottom of the rotationnally broadened stellar lines. Such absorptions could indicate the presence of circumstellar gas, as present in the case of HD39060 \cite[e.g.][]{lagrange00}.

\section{Results} 

\subsection{RV time series and variability properties}

 We show in Appendix A the obtained RV times series for our 26 targets. 
The RV rms and amplitudes, as well as the BVS rms and amplitudes for stars with enough data points, are reported in Table~\ref{obs}. 


The average jitter of our eight A type stars is 237 m/s, while that of the three F-type stars is 35 m/s and that of the thirteen G-K stars is 131 m/s. The median is 67 m/s in the latter case. For the stars for which the BVS could be computed, the average BVS is 61 m/s for the F-type stars and 74 m/s for the G-K type stars. We recall though that within a given spectral type range, there is a large dispersion of jitters due to the star ages (see below). Among the 14 F-K stars with at least two epochs\footnote{We do not consider HD177171 in this section dedicated to the origin of the RV jitter, but we will consider it in the next section to get estimates of detection limits on this type of stars as well.}, only one, HD224228, shows no detetectable level of RV variations, with a RV jitter below 5 m/s. Twelve stars have jitters larger than 5 m/s. Among them, only three have jitters larger than 150 m/s (they will be refered to as "high amplitude variable stars"), five have jitters between 50 and  105 m/s  (they will be refered to as "medium amplitude variable stars") and four have jitters  between 5 and 50 m/s (they will be refered to as "low amplitude variable stars").
\begin{figure}[htp!]
    \centering
	\includegraphics[angle=0,width=\hsize]{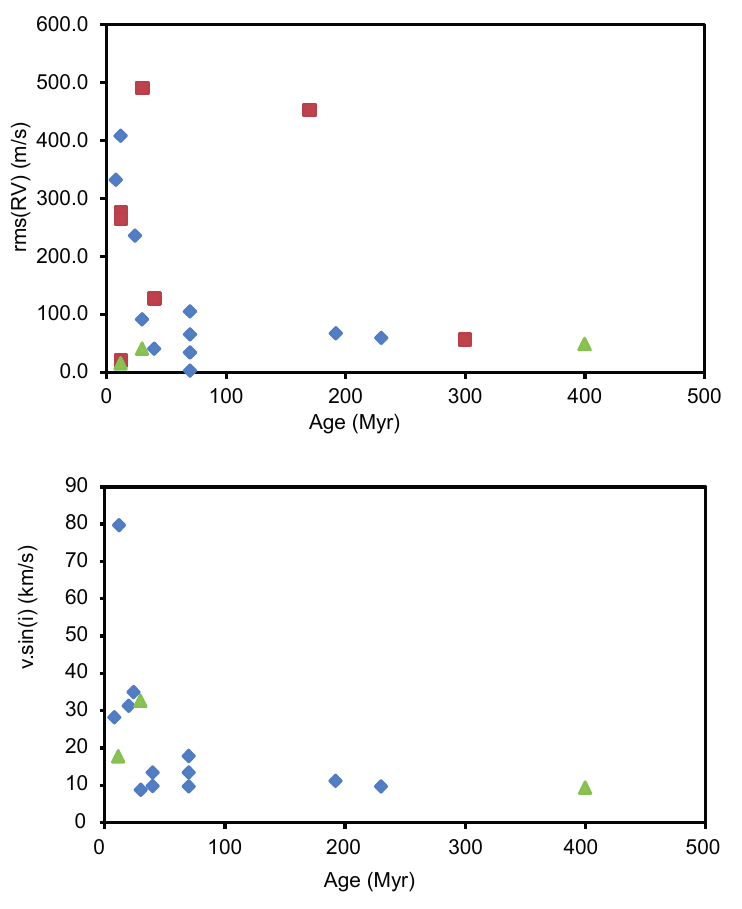}
    \caption{ {\it Upper panel}: jitter versus age for our stars with enough data points (except HD177171), for G-K stars (blue diamonds), F-type stars (green triangles) and A-type stars (red squares). Note that an age of 300 Myr was assumed for Fomalhaut. {\it Bottom panel}: same for \vsini versus age.}
    \label{rms_vsini_age}
  \end{figure}

\begin{figure}[htp!]
    \centering
	\includegraphics[angle=0,width=\hsize]{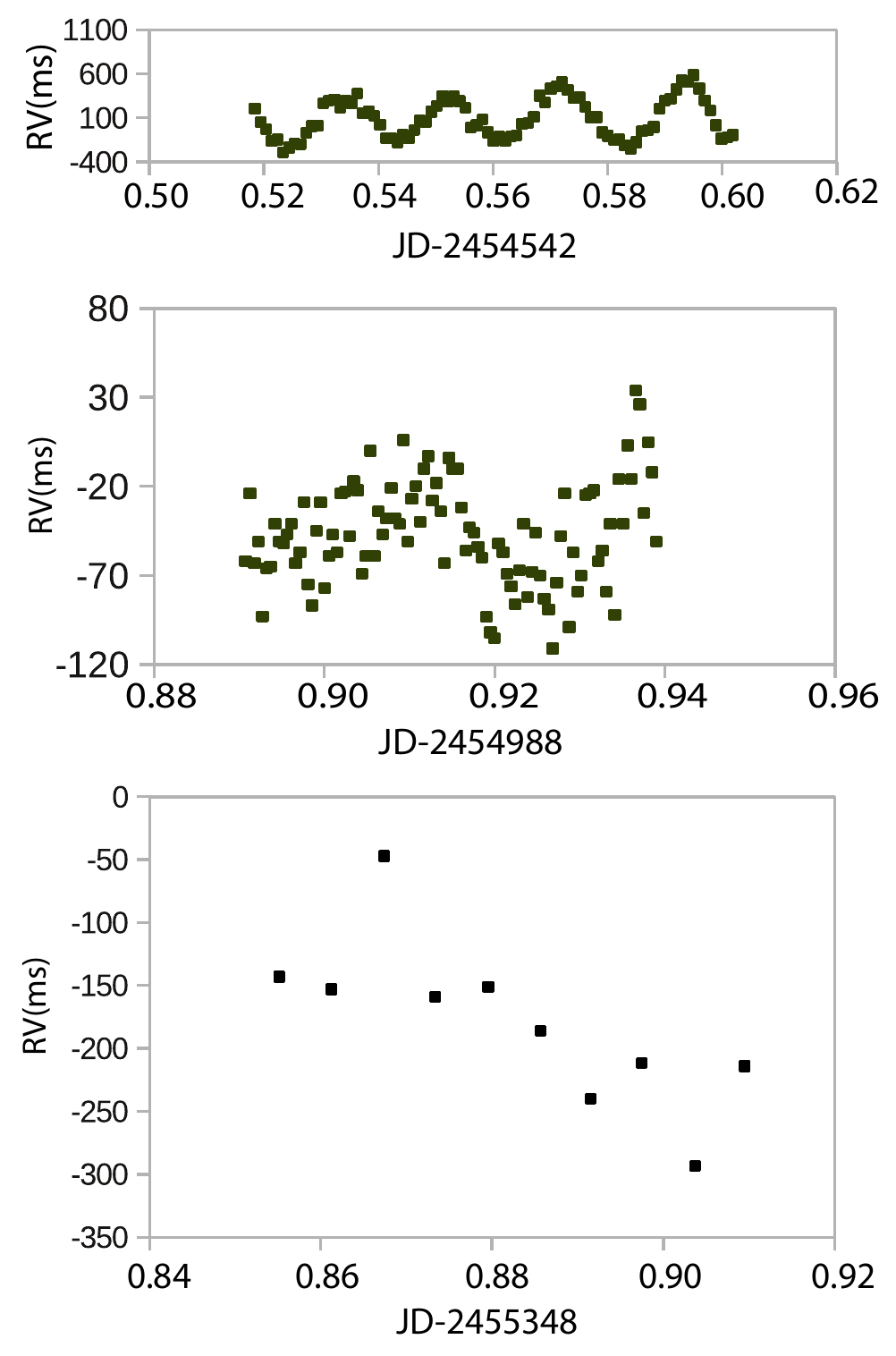}
    \caption{{\it Upper panel}: High frequency variations of RV for HD39060 (exposure time of 90 seconds). {\it Middle panel}: same for HD216956 (exposure time of 8 minutes).  {\it Lower panel}: same for HD218396 (exposure time of 40 seconds).}
    \label{RV_zoom}
  \end{figure}

The \vsini deduced from our cross-correlation functions are also provided in Table~\ref{obs}, whenever possible. For the stars with already published values, our values agree within typically 10 $\%$ with previously reported ones, except for the low \vsini HD224228 for which we find 5.7~km/s instead of 8.7 km/s. 
The average \vsini is 20.1 km/s for F-type stars and 20.5 km/s for G-K stars (with a median of 13.3 km/s). 

We also provide, whenever possible, the R'$_{HK}$ values. They agree within 0.1-0.2 dex with reported ones, although the latter often show discrepancies from one author to the other. The differences can be due either to systematics or to intrinsic stellar variability, which would not be surprising as all these stars (except HD224228) are active. 

Finally, in Fig.~\ref{rms_vsini_age}, we show the RV rms and \vsini as a function of the star ages, as well as the \vsini as a function of RV jitters. We considered only stars with enough data available. We did not consider HD177171 (F6V) either,  because it shows clear signs of being a short period spectroscopic binary, with a possible period of 1.7 days  \cite[we note that HD177171 was classified as a close binary by][on the basis of Hipparcos astrometric data]{frankowski07}. We see that globally, the \vsini of late-type stars significantly decreases with age. This is in qualitative agreement with the results of \cite{weise10} analysis on a larger sample of stars and covering a larger age range. The RV jitter also globally decreases with age, and increases with \vsini. For early-type stars, the correlations between \vsini and age, or RV jitter and age, are not as clear as for late-type stars.

\subsection{Origin of the variability}

We used the bisectors (see Appendix A) as well as  extracted short-term time series of pulsating stars (see examples in Fig.~\ref{RV_zoom}) to classify the observed variations. The results of the classification (activity, pulsations, spectroscopic binarity) are reported in Table~\ref{obs}.

\subsubsection{Early-type stars}

 All our A-type stars show high frequency and high amplitude (up to  5 km/s) RV variations which are characteristic of pulsations (see Table~\ref{obs}). Similar conclusions were reached  in \cite{lagrange09b} for MS stars with similar spectral types.

\subsubsection{Late-type stars}

The RV and BVS of late-type stars are generally well correlated, as shown for example for HD183414 in Fig.~\ref{bsv}, indicating that the variability is due to stellar activity. 
We do not find any correlation between R'$_{HK}$ and RV for any of our stars. This is compatible with the long-term activity of late-type stars beeing dominated by spots instead of plages, as proposed by \cite{lockwood07}. Our observations show a similar effect on shorter timescales.

To help interpreting our results, we ran several cold spots simulations using SAFIR \cite[see a description of the simulations in][]{desort07} to characterize the RV and BVS variations using simulated spectra over the 377-691 nm region. 
 All these simulations are done assuming a solar-type star, but we checked that the results for F5 and G2 stars are quite similar. The spot temperature contrast was 1200 K \cite[][]{berdyugina05}. Also, in these examples, a very high $\mathrm{S/N}$ ($\geq$ 300) was considered, to focus on the effects of star and spot parameters.
Results are provided in  Fig.~\ref{eightshapes1}  in the case of a star seen edge-on with different \vsini, and in 
  Fig.~\ref{eightshapes2}   for different star inclinations, \vsini and spot latitudes.

In most cases (HD37572, HD45270, HD90905, HD102458, HD141493, HD174429, HD181327, HD183414, and HD217343), the (RV, BVS) diagrams indicate activity with a simple pattern, characteristic of either a single spot or slowly evolving spots at the same latitude \cite[][]{desort07}. A more complex case, HD105690, is described below.


We detail here a few cases of interest. 

\begin{itemize}
\item[-] HD61005: only three epochs are available, spread over a few days only. The data reveal variability, with an amplitude larger than 100 m/s. The amount of data is not sufficient to determine any periodicity in these variations, but they are compatible with a 5 day period, as deduced from photometric data by \cite{desidera11}. \cite{setiawan08} observed this star with FEROS and reported strong variations with an amplitude of 150 m/s. Within the 10 m/s FEROS precision on RV, they found a correlation between RV and BVS as an indication of the presence of spots. Our results are compatible with these conclusions.


\item[-] HD105690: the RV and BVS data shown in Fig.~\ref{hd105690} are rather puzzling and very different from those of HD183414 for example (Fig.~\ref{bsv}). Most of the data are compatible with a single spot with a period of 4.88 days, as checked by fitting the RV data by a set of two  signals with periods 4.88 and 2.44 days, which leads to a correlation  between RV and BVS. However, the data recorded between JD 2454343 and JD2454348 lead to a more complex (RV,BVS) diagramme, with an additional, more inclined and off-centered contribution. The RV variations between JD 2454343 and JD2454348 are nonetheless still compatible with a period of about 4.8 days. The comparison with spot simulations shows that the observed (RV, BVS) characteristics can be reproduced only when assuming a small star inclination, of the order of $\simeq$ 10-20 degrees with respect to pole-on. If we do not consider the data taken between JD 2454343 and JD2454348, the observed RV are compatible with a spot with a projected surface of about 1$\%$ at $\simeq$ 30 degrees from the star pole. The spot latitude is not well constrained as the star is close to pole-on. Obviously, this spot must be long-lived. This is in agreement with our knowledge on young stars, even though short-lived spots can sometimes  be found as well, as shown by \cite{garciaalvarez11}. 
                               
We tried to see whether we could reproduce the RV and BVS properties for the {\it whole } period by adding a  short-lived spot. We did not find any configuration that would produce an acceptable match. We can therefore exclude additional spots at a similar latitude as the main, long-lasting spot, or a spot at a lower latitude. 
More data and studies will be needed to understand HD105690 RV variations.

\begin{figure*}[htp!]
    \centering
	\includegraphics[angle=0,width=\hsize]{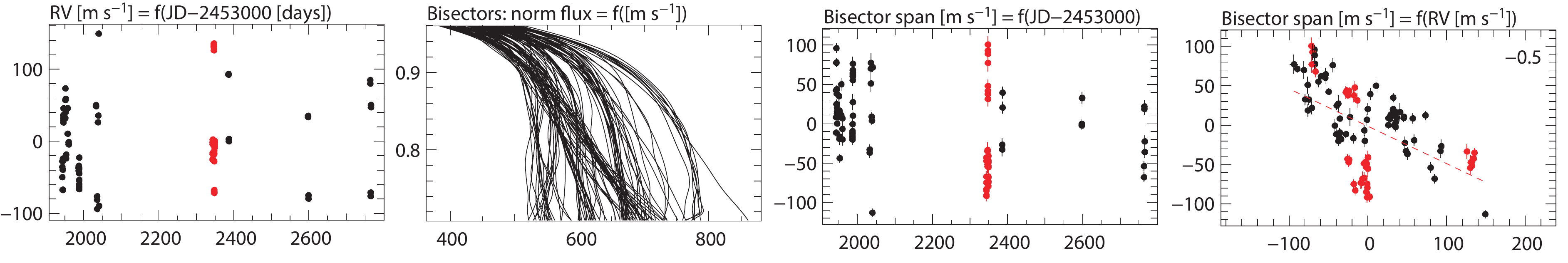}
    \caption{{\it From left to right}:  RV variations of HD105690, bisectors, bisector velocity span (BVS) versus time and BVS versus RV. In the (RV, BVS) diagram, the data corresponding to the period from JD 2454343 to JD2454348 are in red.}
    \label{hd105690}
  \end{figure*}

\item[-] HD174429: it shows high amplitude RV and BVS variations\footnote{Note that the available data do not allow to detect the trend due to PZ Tel B, due to the limited time span available, and the amplitude of the RV variations.} (see Fig.~\ref{hd174429_bsv}). The observed bisector shape is closer to that of pulsating stars than to that of low \vsini late-type active stars. 
Simulations show that for a given spot size, the RV and the BSV amplitudes significantly increase with \vsini (see Fig.~\ref{eightshapes1} and Fig.~\ref{eightshapes2}, so we conclude that the high RV and BVS amplitudes can be just a consequence of the star high \vsini. Note that the PZ Tel \vsini, 80 km/s, is the highest projected rotational velocity among our late-type stars. \cite{garciaalvarez11} proposed that this  high \vsini is due to accretion from a close TT star. We do not see spectroscopic signs of accretion, though. 

\begin{figure*}[htp!]
    \centering
	\includegraphics[angle=0,width=\hsize]{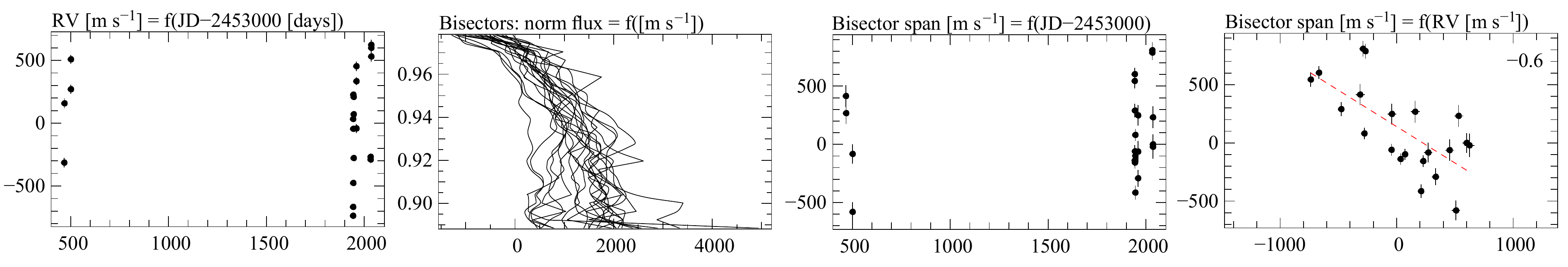}
    \caption{{\it From left to right}:  RV variations of HD174429, bisectors, bisector velocity span (BVS) versus time and BVS versus RV.}
    \label{hd174429_bsv}

  \end{figure*}

\item[-] HD207575: its (RV, BVS) shape is quite different from that of our other active F stars, and closer to that of earlier type stars. However, for a given spot pattern, the RV and BVS characteristics strongly depend on \vsini, as illustrated in Fig.~\ref{eightshapes1} and Fig.~\ref{eightshapes2}. The (RV, BVS) shape departs from the roughly linear or even from the "eight-shape" form for stars seen with very low (5 degrees) inclinations from pole-on and with \vsini in the range 30-40 km/s. Such a situation is however not compatible with the star spectral type and observed \vsini of 35 km/s.

\end{itemize}
\subsection{Detection limits}

\subsubsection{Obtained detection limits}

 In Appendix A, we provide  the detection limits computed with the rms-based  method and the LPA ones, when computed. The corresponding LPA-derived values for periods of 3, 10, 100, 500 and 1000 days are reported in Table~\ref{obs}. We recall that the LPA method provides more accurate and better detection limits than the rms-based method. The rms-based limits are provided here only because for several stars, we do not have enough data to compute the LPA detection limits. 
In Appendix A, we also indicate the rms-based detection limits that would  theoretically be achieved with a much better temporal sampling ("achievable rms-based detection limits"), as well as the rms-based detection limits that would be achieved if the stars were not active or pulsating, i.e. if the noise was dominated by photon/instrumental noises.

The detection limits depend on both the monitoring quality (time-span and sampling) and on the stars (spectral-type, \vsini and activity/pulsations). For most stars, the time-spans of the RV monitoring allow us to compute significant detection limits for periods smaller than a few hundreds days. For a few targets, the time-span is close to or larger than 1000 days. This is the case for HD39060, HD71555, HD105690, HD146624, HD172555, HD174429, HD188228, HD216956, and HD218396. 

Globally, the {\it conservative } rms-based achievable limits are all below 10 \Mjup for periods less than 100 days, except for the A-type highly pulsating stars HD71555 and HD218396. As expected, the LPA method provides detection limits better than the rms-based limits for all our stars.  

For early-type stars, the detection limits strongly depend on the available data. This is very clear if one compares HD39060 (\bp) and HD172555, which have similar spectral types and RV jitters. The detection limits can be substantially improved when averaging RV obtained on long series of data (1-2 hours). Planets of 2-3 \Mjup can be found at periods up to 100 days on HD39060 \cite[see an extensive discussion in][]{lagrange12}, to be compared to masses of 4-10 \Mjup for similar periods for HD172555.  HD109536, with a temporal sampling covering 93 days only, has detection limits in the range 1-2 \Mjup for periods up to tens of days, even though its rms(RV) is as large as 216~m/s. {\it We conclude that provided enough and well sampled data are available, it is possible to detect giant planets down to a few Jupiter masses, even with long periods, around these young early-type stars. }

Note that excellent detection limits are obtained for HD216956 (Fomalhaut) for periods up to more than 1000 days (Fig.~\ref{hd216956_average}). A $\simeq$ 30 \Mjup brown dwarf at 5 AU was proposed by \cite{chiang09} to explain the observed Hipparcos acceleration. The present detection limits  do not allow to rule out such a companion, because the time-span is not long enough, but the presence of such a companion could easily be tested provided a longer time-span and an appropriate observing strategy, as we did in the case of \bp \cite[][]{lagrange12}. 

Detection limits for late-type stars for periods below 100 days are most of the time lower than 1 \Mjup. Only for the few stars with the highest RV jitters, for example for HD141593 which has a RV jitter of 235 m/s, these detection limits become of the order of 5.5 \Mjup for periods of 100 days. Stars with intermediate RV jitters, which are the most representative of our sample, have much lower detection limits. For the well monitored HD105690, with a RV rms of 60 m/s, we obtain a detection limit of 1.2 \Mjup for a period of 100 days, and of 2 \Mjup for periods around 800 days. HD183414, with RV rms of 60 m/s, has sub-Jupiter mass limits up to 100 days. Finally, HD181327, which has a lower RV jitter of 15 m/s, has a detection limit between less than 0.1 \Mjup and 0.2 \Mjup for periods up to 40 days, and 0.3 \Mjup for a period of 100 days. Note that the  periods considered here are limited only by the time spans of the data available. Additional data would allow to increase the range of periods to be considered.

{\it We conclude that when enough and well-sampled data are available, it is possible to find giant planets down to very low masses, down to 1 \Mjup or below, orbiting these active young late-type stars, up to long periods. }



\subsubsection{Improvements of the detection limits}

\begin{figure*}[htp!]
    \centering
	\includegraphics[angle=0,width=\hsize]{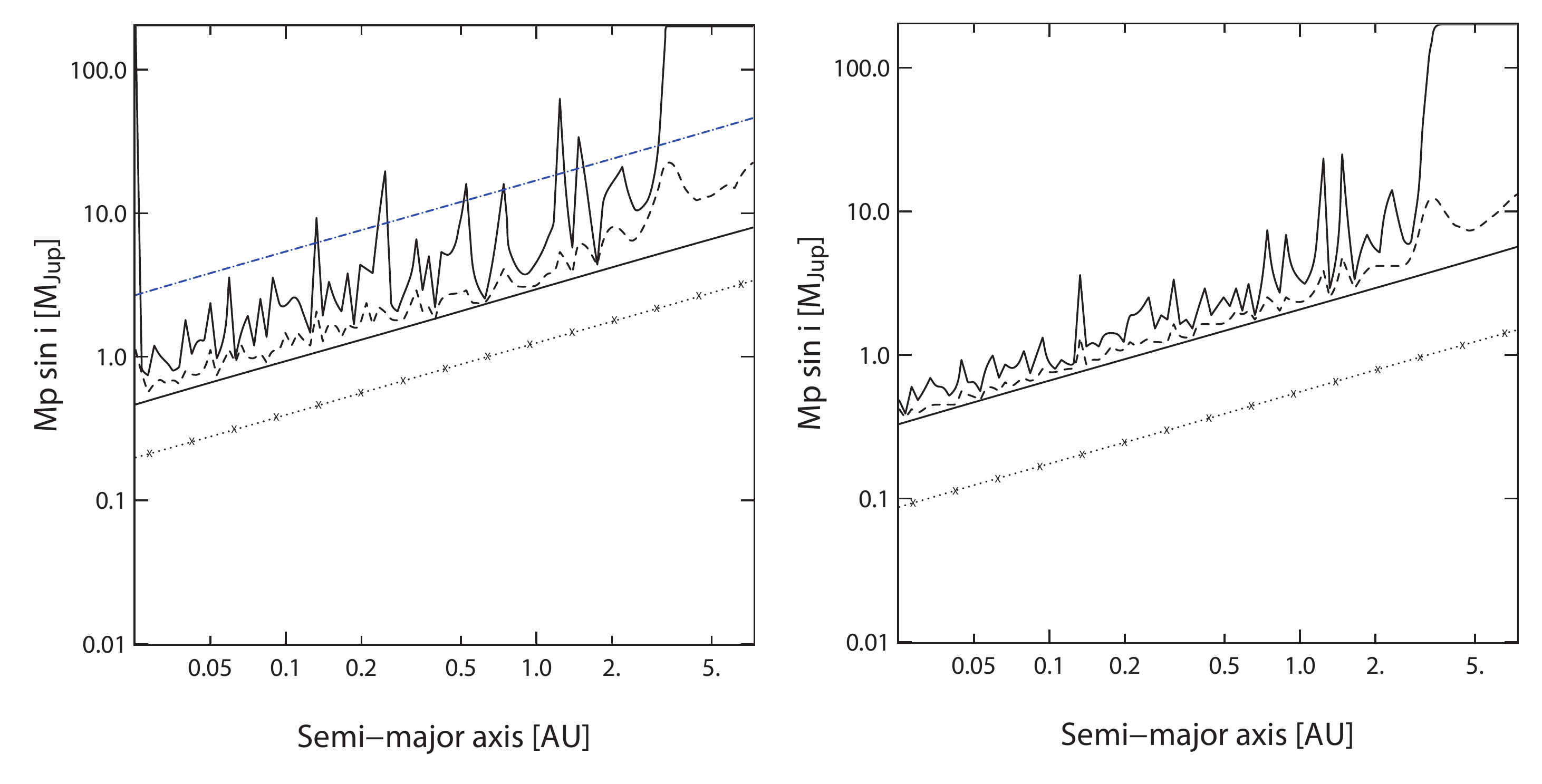}
    \caption{Impact of data averaging on pulsating stars: rms-based detection limits obtained on HD 216956 when using all measurements directly ({\it left panel}), and when averaging the data over 2 days ({\it right panel}). Same symbols as for the middle panels of Fig.\ref{rv_limdet_1}.}
    \label{hd216956_average}
  \end{figure*}

As mentioned above, adopting  adequate observing and averaging strategies is particularly efficient to improve the detection limits for early-type, pulsating stars. This is also illustrated in Fig.~\ref{hd216956_average} in the case of HD216956, for which we recorded at each epoch long data sets of 1-2 hours. In this case, the RV jitter is reduced from 56~m/s to 25~m/s after averaging. 

For active stars, the detection limits obtained when taking into account the temporal structure of the stellar noise are already very satisfactory. Nonetheless, it is worth trying to improve our detection capabilities. Several approaches have been proposed and gave more or less successful results on solar-type MS stars: use of simultaneous photometric variations to correct from spots/plages \cite[][and references therein]{aigrain12,lanza11}, use of the BVS \cite[][and references therein]{boisse11} and use of spectroscopic activity indicators \cite[][]{boisse11}. Using simultaneous photometric data may be quite powerful when the signal is not dominated by convection \cite[see an analysis in][]{meunier13}. Using the BVS or other spectroscopic diagnostics is very tempting as these parameters are obtained exactly simulatenously with the RV data and do not request additional observations. 

As seen previously, we do not observe any correlation between the Ca emission and the RV variations, so we cannot use the Ca emissions to improve our detection limits.
We then consider the use of the BVS. The \vsini of our stars are usu ally significantly higher than the spectral resolution, and this induces complex relationships between the BVS and the RV, as demonstrated by \cite{desort07} and \cite{boisse11}. This is also seen in Appendix B 
 in a number of examples. Departures from a linear correlation such as inclined "eightshapes" or even more complex structures can be observed. The amount of departure from a simple linear dependance depends on the star \vsini, on the star orientation with respect to the line of sight $\sin{i}$ and on the spots parameters.  

\begin{figure}[htp!]
    \centering
\includegraphics[angle=0,width=\hsize]{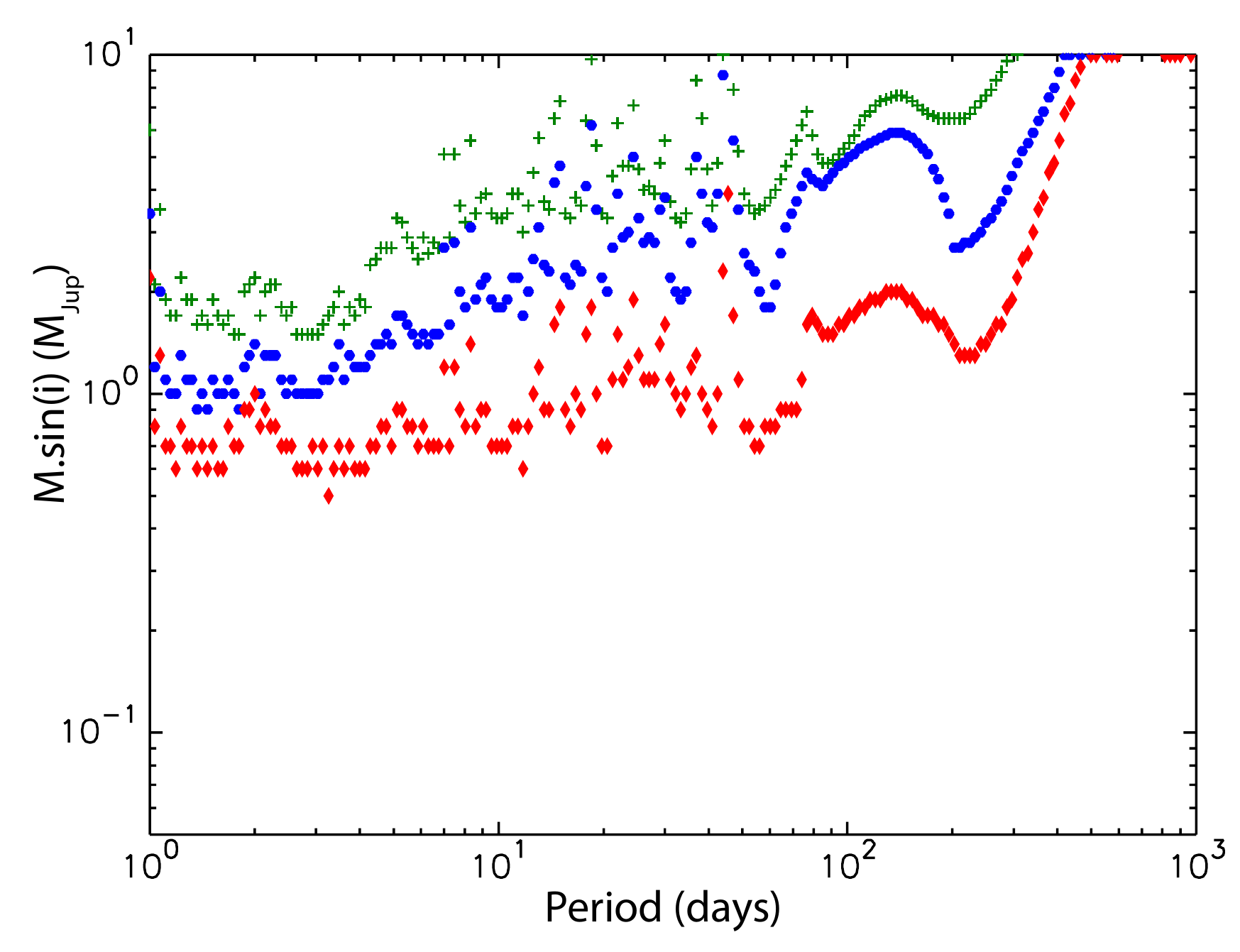}\\
\includegraphics[angle=0,width=\hsize]{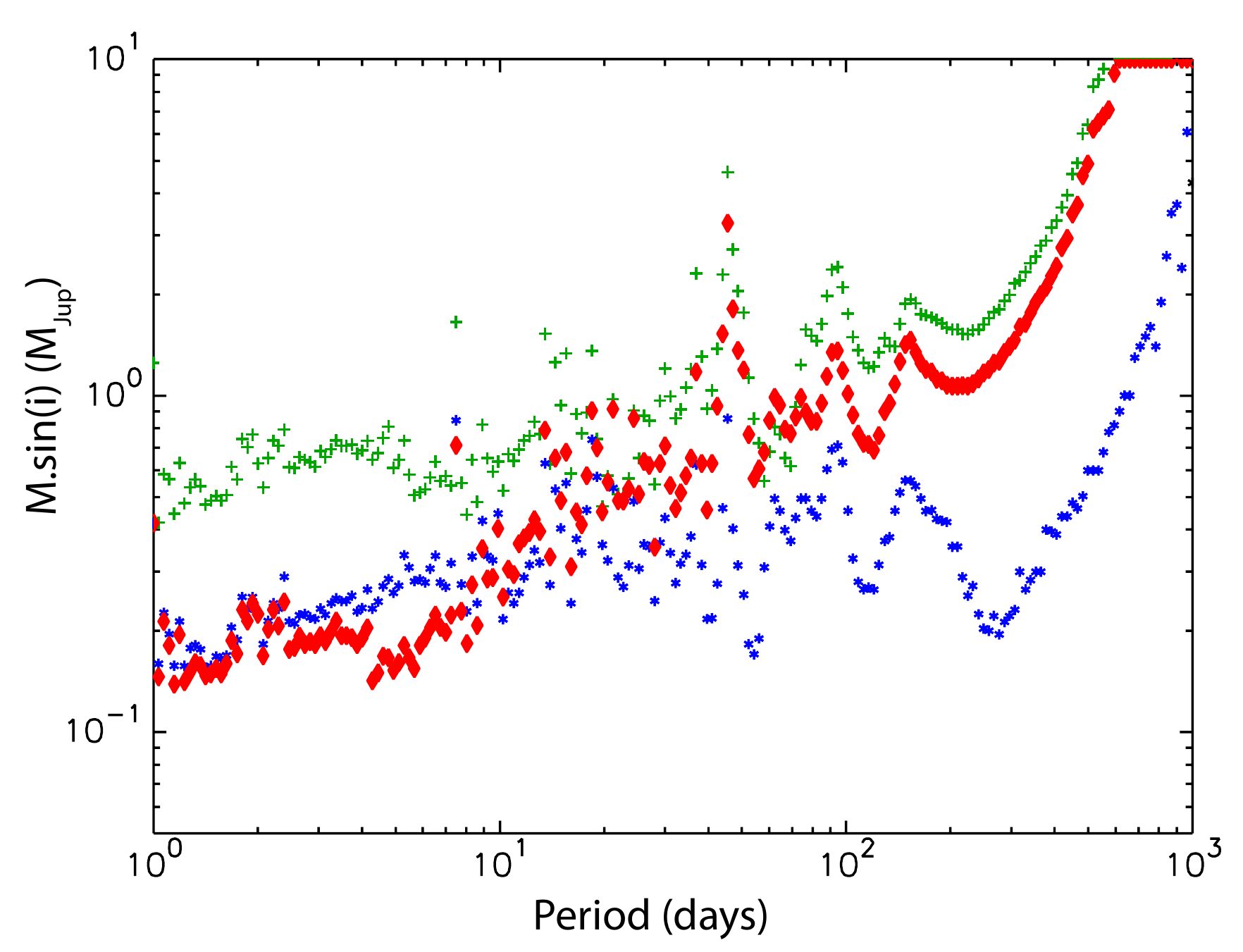}\\
\includegraphics[angle=0,width=\hsize]{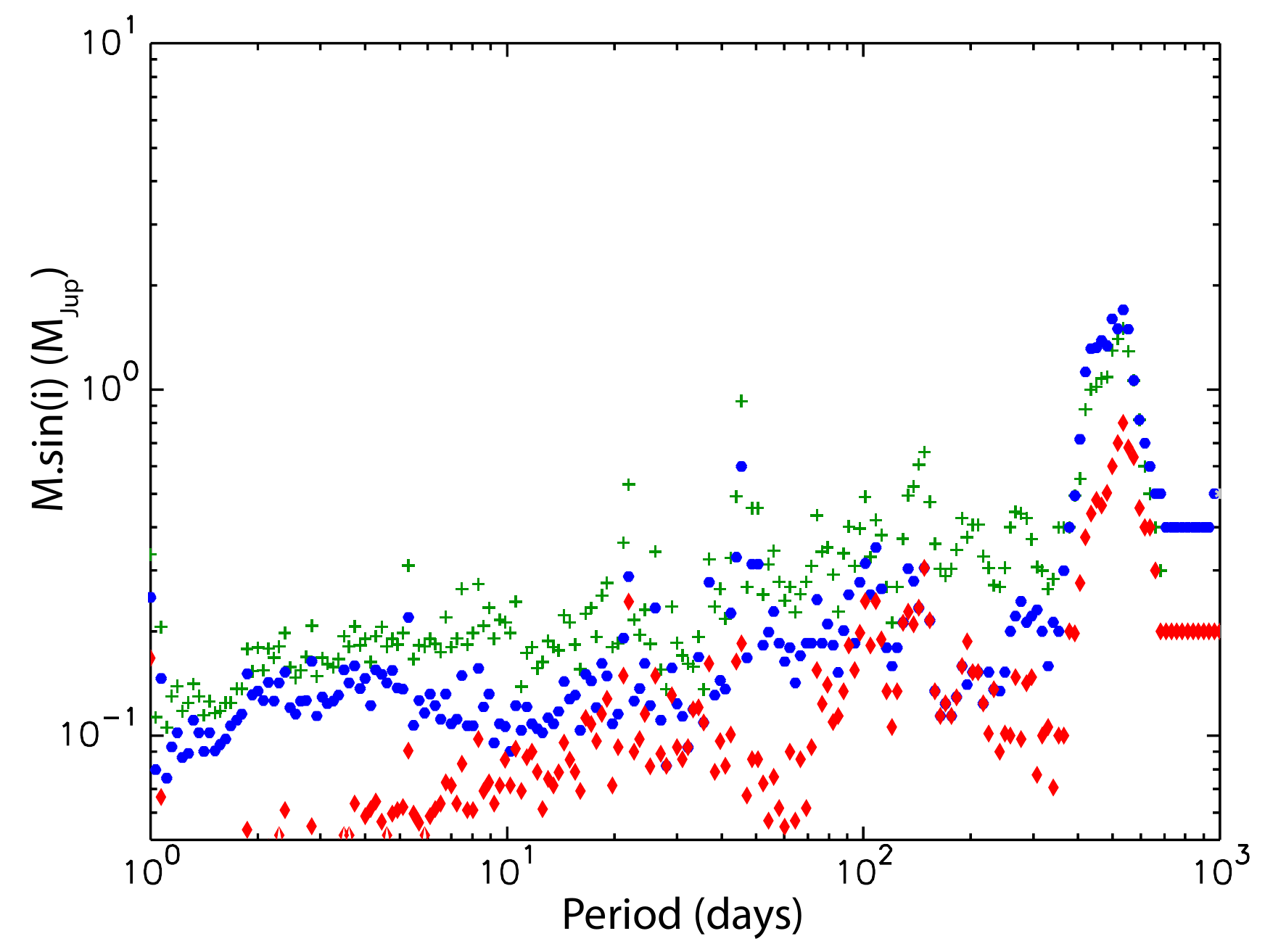} \\
    \caption{{\it Upper panel}: LPA detection limits after correction of the RV using the (BVS,RV) slope (blue dots), and after fitting and removing the spot signal (red diamonds), for HD141943 (jitter = 235 m/s, \vsini = 35 km/s). The uncorrected LPA detection limits are indicated by green crosses. {\it Middle panel}: same for HD183414 (jitter = 67 m/s, \vsini = 11.1 km/s). {\it Lower panel}: same for HD181327 (jitter = 15 m/s, \vsini = 17.9 km/s).}
    \label{limdet_corr}
  \end{figure}

We focused on the three active stars with the largest number of data and for which a good (RV, BVS) correlation trend (hereafter slope) could be determined, namely (with the measured RV jitter and \vsini in parenthesis): HD141943 (235 m/s, 35 km/s), HD181327 (15 m/s, 17.9 km/s) and HD183414 (67 m/s, 11.1 km/s). The slopes of the (RV, BVS) correlation are indicated in Table~\ref{obs}. 
After correction of the RV using the (BVS, RV) correlation, the RV jitters are significantly reduced, as well as the detection limits (see Fig.~\ref{limdet_corr}). For HD141943, the RV jitter of the corrected RV is reduced from 238 m/s to 131 m/s, and the detection limits are reduced by a factor of 1.6 on average; they now fall in the range 1-5 \Mjup for periods below 100 days. For HD183414, the RV jitter is reduced from 67~m/s to 25 m/s, and the detection limits are reduced by a factor larger than 2.9 on average; they are at the Saturn mass level for periods below 100 days. Finally, for HD181327, the RV jitter is reduced from 15~m/s to 12 m/s, and the detection limits are reduced by a factor of 1.5; they are in the 2-4 Neptune mass domain for periods below  100 days.


An even better correction could be expected by removing the spot signal itself, using for example simultaneous photometric data. A perfect correction would remove all activity-induced RV signal, and the achievable detections limits would be those achievable under photon and instrumental noises only (see the third panels of 
Fig.~\ref{rv_limdet_1}). As an exercice and for illustration purposes only, we tried to fit the observed RV of the same stars with a keplerian signal (mimicking a spot) and then estimate the detection limits on the residuals. Significant improvements were obtained for HD141943 and HD181327, as seen in Fig.~\ref{limdet_corr}: the detection limits after correction now reach levels below 1 \Mjup for the very active star HD141943 and of 1-2 Neptune mass planets for the weakly active star HD181327. However, this approach assumes that the spot signal can be fitted by a  keplerian signal, which is not always the case. In particular, this assumption does not apply when the spot is not always visible or when the activity pattern is complex, as illustrated by the simulations showed above. 
Hence, even though the detection limits are significantly improved, we would not advise to use this method for planet detection unless simultaneous photometric data are available to provide independant constraints on the spot(s) properties.

\subsubsection{Comparison with previous surveys}
Before comparing our detection limits to those resulting from other works, we recall that these quantities strongly depend on the targets themselves (jitter and temporal distribution of the stellar noise), on the observing strategy, and also on the way the limits are measured, as extensively discussed in \cite{meunier12}. Our detection limits are much better than those observed by \cite{bailey12}. However, the later focused on  stars with spectral types M, i.e. later than those of our targets, and their detection limit  estimation is probably conservative (like our rms-based method). Our detection limits, even before correcting from activity, are better than those obtained by \cite{paulson06} on the $\beta$ Pic moving group; for these targets, their detection limits at short periods fall in the brown dwarf regime, where as our limits are well in the planet regime for similar periods.  We attribute this to the internal precision of their measurements, their adopted observing strategy, and their approach to estimate the detection limits. Finally, the fact that we find much lower detection limits for older stars than for younger ones is in agreement with \cite{paulson04} results.

\subsection{Search for circumstellar Ca in rapid rotators}

 Finally, in no cases did we find evidence of circumstellar lines, except in the case of \bp. In the case of the A-type \bp analogous star HD172555, a narrow and faint component is present at the bottom of the Ca line. However, it is most probably of interstellar origine.

\section{Towards a full exploration of young stars giant planet population}

We have shown that RV monitoring with an appropriate observing strategy and data analysis can allow to detect giant planets around young nearby stars despite their activity. The separations of detectable planets with RV techniques are however limited to a few AU. To get a complete view of planetary systems, it is  necessary to complement RV data with other types of data: astrometric ones, e.g. Gaia, for typical separations of 2-5 AU \cite[][]{sozzetti11}, and, above all, direct imaging data which can provide detection limits at much larger separations. Direct imaging is particularly interesting for young systems as the contrasts between young planets and their parent stars are more favorable than for older systems. The combination of these techniques therefore offers a unique opportunity to
explore the giant planet population on a same class of stars. We illustrate in this section the potential of the coupling of RV and AO deep imaging data, using the example of Harps and NaCo, and Sphere data. 

\begin{figure*}[htp!]
    \centering
\includegraphics[angle=0,width=1.\hsize]{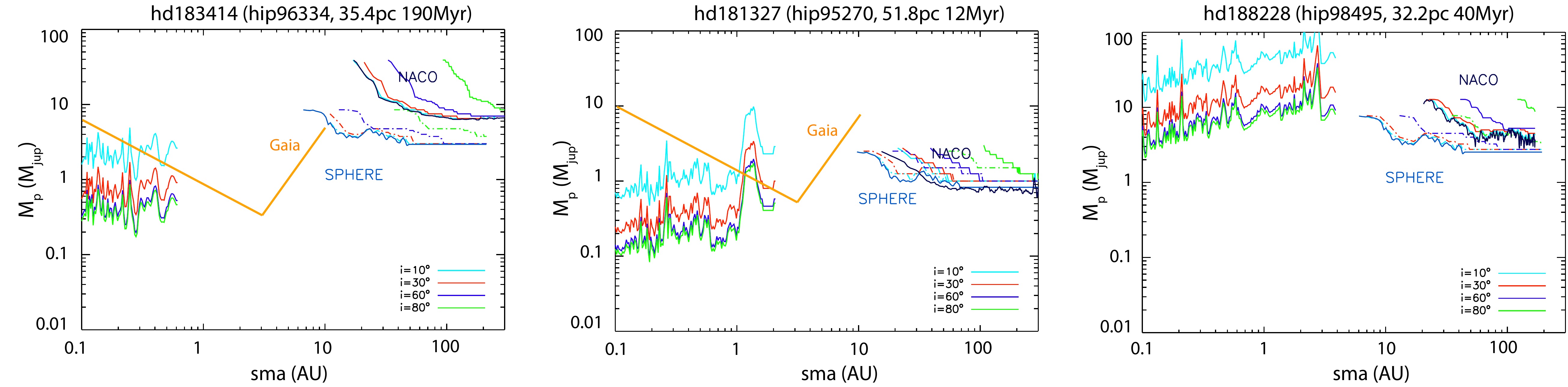}
    \caption{ {\it Upper panel}: AO detection limits (observed NaCo limits in plain lines and expected SPHERE limits in dotted lines) and Harps RV detection limits for the active G-type star HD183414 (192 Myr, 35 pc and time-span of 189 days). In both cases, different star inclinations: 10 (light blue), 30 (red), 60 (blue) and 80 degrees (green) are considered. The Gaia detection limits are also indicated. 
{\it Middle panel}: idem for the less active F6 star HD181327 (12 Myr, 51 pc and time-span of 588 days).
{\it Lower panel}: same for the pulsating A-type star HD188228 (40 Myr, 32 pc and  time-span of 1984 days). Note that as the star is brighter than V=6, no Gaia data will be available for this star.}
    \label{couplage_AO_RV}
  \end{figure*}

We consider three stars of our sample for which we have both Harps RV and NaCo imaging data, and which could be considered as prototypes for such complementary studies, namely HD183414, HD181327, and HD188228. The G-type star HD183414 is an example of medium activity-level star, with a jitter of 67m/s; it is located 35 pc away. The F-type star HD181327, located at 51 pc, has a low activity level (jitter=15 m/s). The pulsating A-type star HD188228, located at 32 pc, has a jitter of 127 m/s (amplitude of 2 km/s). We plot in Fig.~\ref{couplage_AO_RV} the RV and imaging detection limits expressed in Jupiter masses for these three stars. The deep imaging detection limits are either those measured on NaCo data \cite[][]{rameau13}, and the ones expected with Sphere \cite[][]{mesa11}. Note that the detection limits of HD183414 are those obtained once correction for activity using the RV-BVS relation has been made. Different star inclinations are considered for the RV data. The limits are strongly impacted for inclinations below 60 degrees. For larger inclinations, the RV derived detection limits are very similar. 
 We  see that the detection limits are in the planetary regime with both approaches. In these examples, the RV limits are much better than the NaCo ones for the G-type stars (this effect is even stronger when considering moderately active stars with a RV jitter below 50 m/s), while the AO limits are better than the RV ones for the A-type star. 

The complementarity of HARPS and NaCo in terms of separation ranges is obvious. However, gaps are still present. This is particularly true for HD183414, for which the available RV data are spread over 189 days only.  On the contrary, for HD188288, which is located at a similar distance, there is an excellent overlap because we dispose of RV data spread over about 2000 days. Finally, the gap observed for HD181327 is smaller than that of HD183414 because the available RV data span is much larger (560 days), even though the star is located further away. We note the important impact of forthcoming imagers like 
SPHERE on the VLT: with inner working angles smaller than 0.15"-0.2", they will allow the detection of planets much closer to the stars than current imagers such as NaCo; this will fill the gap between RV data and imaging ones for stars closer than typically 30 pc; this will also significantly contribute to reduce the gap for stars located further away. Finally, we see that Gaia can help filling the gap, if any, between RV and AO direct imaging data. This will be particularly important  for distant stars.

\section{Concluding remarks}
We have shown that RV data allow the detection of giant planets around stars in close-by, young associations. Given the available data, we could explore different separations ranges, up to more than 2 AU for some stars of our sample. For low to moderately active stars, detection limits of a few Neptune masses are obtained. The best detection limits are obtained for low to moderate \vsini in the range 10-30 km/s, which are generally observed for stars older than 30 Myr, although there are some exceptions. Early-type stars are generally pulsating, and averaging over the high frequency variations reduces considerably the achievable detection limits. Values well within the giant planet domain are obtained.

 Finally, we have illustrated the tremendous interest of coupling RV and deep AO data using Harps and NaCo data, and, even more, future Sphere data. RV data covering more than 2000 days allow planet detection and characterization for separations up to 3-5 AU, while forthcoming imagers will provide excellent detection limits for separations greater than 0.15-0.2", which correspond to 4-5 AU at 30 pc. For more distant stars, Gaia will fill the remaining gap. We will have in the next decade, and for the first time, a  unique opportunity  to make an exhaustive giant planet search around these stars, that will in turn lead to unprecedented studies of giant planet formation and evolution.

\begin{acknowledgements}

  We acknowledge support from the French CNRS and the support from the Agence Nationale de la Recherche (ANRBLANC10-0504.01). We are grateful to the ESO staff for their help during the observations, and to the Programme National de Plan\'etologie (PNP, INSU).  These results have made use of the SIMBAD astronomical database, operated at the Centre de Données Astronomiques de Strasbourg,  France, and NASA’s Astrophysics Data System.  We also thank G\'erard Zins and Sylvain C\`etre for their help in implementing the SAFIR interface. 
\end{acknowledgements}

\bibliographystyle{aa}
\bibliography{biblio}

\begin{thebibliography}{64}
\expandafter\ifx\csname natexlab\endcsname\relax\def\natexlab#1{#1}\fi

\bibitem[{{Aigrain} {et~al.}(2012){Aigrain}, {Pont}, \& {Zucker}}]{aigrain12}
{Aigrain}, S., {Pont}, F., \& {Zucker}, S. 2012, \mnras, 419, 3147

\bibitem[{{Bailey} {et~al.}(2012){Bailey}, {White}, {Blake}, {Charbonneau},
  {Barman}, {Tanner}, \& {Torres}}]{bailey12}
{Bailey}, III, J.~I., {White}, R.~J., {Blake}, C.~H., {et~al.} 2012, \apj, 749,
  16

\bibitem[{{Berdyugina}(2005)}]{berdyugina05}
{Berdyugina}, S.~V. 2005, Living Reviews in Solar Physics, 2, 8

\bibitem[{{Beuzit} {et~al.}(2008){Beuzit}, {Feldt}, {Dohlen}, {Mouillet},
  {Puget}, {Wildi}, {Abe}, {Antichi}, {Baruffolo}, {Baudoz}, {Boccaletti},
  {Carbillet}, {Charton}, {Claudi}, {Downing}, {Fabron}, {Feautrier},
  {Fedrigo}, {Fusco}, {Gach}, {Gratton}, {Henning}, {Hubin}, {Joos}, {Kasper},
  {Langlois}, {Lenzen}, {Moutou}, {Pavlov}, {Petit}, {Pragt}, {Rabou}, {Rigal},
  {Roelfsema}, {Rousset}, {Saisse}, {Schmid}, {Stadler}, {Thalmann}, {Turatto},
  {Udry}, {Vakili}, \& {Waters}}]{beuzit08}
{Beuzit}, J.-L., {Feldt}, M., {Dohlen}, K., {et~al.} 2008, in Society of
  Photo-Optical Instrumentation Engineers (SPIE) Conference Series, Vol. 7014

\bibitem[{{Biller} {et~al.}(2010){Biller}, {Liu}, {Wahhaj}, {Nielsen}, {Close},
  {Dupuy}, {Hayward}, {Burrows}, {Chun}, {Ftaclas}, {Clarke}, {Hartung},
  {Males}, {Reid}, {Shkolnik}, {Skemer}, {Tecza}, {Thatte}, {Alencar},
  {Artymowicz}, {Boss}, {de Gouveia Dal Pino}, {Gregorio-Hetem}, {Ida},
  {Kuchner}, {Lin}, \& {Toomey}}]{biller10}
{Biller}, B.~A., {Liu}, M.~C., {Wahhaj}, Z., {et~al.} 2010, \apjl, 720, L82

\bibitem[{{Boisse} {et~al.}(2011){Boisse}, {Bouchy}, {H{\'e}brard}, {Bonfils},
  {Santos}, \& {Vauclair}}]{boisse11}
{Boisse}, I., {Bouchy}, F., {H{\'e}brard}, G., {et~al.} 2011, \aap, 528, A4

\bibitem[{{Boisse} {et~al.}(2009){Boisse}, {Moutou}, {Vidal-Madjar}, {Bouchy},
  {Pont}, {H{\'e}brard}, {Bonfils}, {Croll}, {Delfosse}, {Desort}, {Forveille},
  {Lagrange}, {Loeillet}, {Lovis}, {Matthews}, {Mayor}, {Pepe}, {Perrier},
  {Queloz}, {Rowe}, {Santos}, {S{\'e}gransan}, \& {Udry}}]{boisse09}
{Boisse}, I., {Moutou}, C., {Vidal-Madjar}, A., {et~al.} 2009, \aap, 495, 959

\bibitem[{{Borgniet} {et~al.}(2013){Borgniet}, {Lagrange}, {Boisse}, {Bouchy},
  { Arnold}, {Diaz}, {Santerne}, {Hebrard}, {Galland}, {Delorme}, {Ehrenreich},
  {Segransan}, { Bonfils}, {Delfosse}, X., {Santos}, { Forveille}, {Moutou},
  {Udry}, {Eggenberger}, \& { Pepe}}]{borgniet13}
{Borgniet}, S., {Lagrange}, A.-M., {Boisse}, I., {et~al.} 2013, submitted to
  A\&A, 1, 1

\bibitem[{{Buenzli} {et~al.}(2010){Buenzli}, {Thalmann}, {Vigan}, {Boccaletti},
  {Chauvin}, {Augereau}, {Meyer}, {M{\'e}nard}, {Desidera}, {Messina},
  {Henning}, {Carson}, {Montagnier}, {Beuzit}, {Bonavita}, {Eggenberger},
  {Lagrange}, {Mesa}, {Mouillet}, \& {Quanz}}]{buenzli10}
{Buenzli}, E., {Thalmann}, C., {Vigan}, A., {et~al.} 2010, \aap, 524, L1

\bibitem[{{Chelli}(2000)}]{chelli00}
{Chelli}, A. 2000, \aap, 358, L59

\bibitem[{{Chiang} {et~al.}(2009){Chiang}, {Kite}, {Kalas}, {Graham}, \&
  {Clampin}}]{chiang09}
{Chiang}, E., {Kite}, E., {Kalas}, P., {Graham}, J.~R., \& {Clampin}, M. 2009,
  \apj, 693, 734

\bibitem[{{Crockett} {et~al.}(2012){Crockett}, {Mahmud}, {Prato},
  {Johns-Krull}, {Jaffe}, {Hartigan}, \& {Beichman}}]{Crockett12}
{Crockett}, C.~J., {Mahmud}, N.~I., {Prato}, L., {et~al.} 2012, \apj, 761, 164

\bibitem[{{Desidera} {et~al.}(2011){Desidera}, {Covino}, {Messina}, {D'Orazi
  D'Orazi}, {Alcal{\'a}}, {Brugaletta}, {Carson}, {Lanzafame}, \&
  {Launhardt}}]{desidera11}
{Desidera}, S., {Covino}, E., {Messina}, S., {et~al.} 2011, \aap, 529, A54

\bibitem[{{Desort} {et~al.}(2007){Desort}, {Lagrange}, {Galland}, {Udry}, \&
  {Mayor}}]{desort07}
{Desort}, M., {Lagrange}, A.-M., {Galland}, F., {Udry}, S., \& {Mayor}, M.
  2007, \aap, 473, 983

\bibitem[{{Dumusque} {et~al.}(2011){Dumusque}, {Lovis}, {S{\'e}gransan},
  {Mayor}, {Udry}, {Benz}, {Bouchy}, {Lo Curto}, {Mordasini}, {Pepe}, {Queloz},
  {Santos}, \& {Naef}}]{dumusque11}
{Dumusque}, X., {Lovis}, C., {S{\'e}gransan}, D., {et~al.} 2011, \aap, 535, A55

\bibitem[{{Duncan} {et~al.}(1991){Duncan}, {Vaughan}, {Wilson}, {Preston},
  {Frazer}, {Lanning}, {Misch}, {Mueller}, {Soyumer}, {Woodard}, {Baliunas},
  {Noyes}, {Hartmann}, {Porter}, {Zwaan}, {Middelkoop}, {Rutten}, \&
  {Mihalas}}]{duncan91}
{Duncan}, D.~K., {Vaughan}, A.~H., {Wilson}, O.~C., {et~al.} 1991, \apjs, 76,
  383

\bibitem[{{Figueira} {et~al.}(2010){Figueira}, {Marmier}, {Bonfils}, {di
  Folco}, {Udry}, {Santos}, {Lovis}, {M{\'e}gevand}, {Melo}, {Pepe}, {Queloz},
  {S{\'e}gransan}, {Triaud}, \& {Viana Almeida}}]{figueira10}
{Figueira}, P., {Marmier}, M., {Bonfils}, X., {et~al.} 2010, \aap, 513, L8

\bibitem[{{Frankowski} {et~al.}(2007){Frankowski}, {Jancart}, \&
  {Jorissen}}]{frankowski07}
{Frankowski}, A., {Jancart}, S., \& {Jorissen}, A. 2007, \aap, 464, 377

\bibitem[{{Galland} {et~al.}(2005){Galland}, {Lagrange}, {Udry}, {Chelli},
  {Pepe}, {Queloz}, {Beuzit}, \& {Mayor}}]{galland05}
{Galland}, F., {Lagrange}, A.-M., {Udry}, S., {et~al.} 2005, \aap, 443, 337

\bibitem[{{Garc{\'{\i}}a-Alvarez} {et~al.}(2011){Garc{\'{\i}}a-Alvarez},
  {Lanza}, {Messina}, {Drake}, {van Wyk}, {Shobbrook}, {Butler}, {Kilkenny},
  {Doyle}, \& {Kashyap}}]{garciaalvarez11}
{Garc{\'{\i}}a-Alvarez}, D., {Lanza}, A.~F., {Messina}, S., {et~al.} 2011,
  \aap, 533, A30

\bibitem[{{Hern{\'a}n-Obispo} {et~al.}(2010){Hern{\'a}n-Obispo},
  {G{\'a}lvez-Ortiz}, {Anglada-Escud{\'e}}, {Kane}, {Barnes}, {de Castro}, \&
  {Cornide}}]{hernan10}
{Hern{\'a}n-Obispo}, M., {G{\'a}lvez-Ortiz}, M.~C., {Anglada-Escud{\'e}}, G.,
  {et~al.} 2010, \aap, 512, A45

\bibitem[{{Hines} {et~al.}(2007){Hines}, {Schneider}, {Hollenbach}, {Mamajek},
  {Hillenbrand}, {Metchev}, {Meyer}, {Carpenter}, {Moro-Mart{\'{\i}}n},
  {Silverstone}, {Kim}, {Henning}, {Bouwman}, \& {Wolf}}]{hines07}
{Hines}, D.~C., {Schneider}, G., {Hollenbach}, D., {et~al.} 2007, \apjl, 671,
  L165

\bibitem[{{Isaacson} \& {Fischer}(2010)}]{isaacson10}
{Isaacson}, H. \& {Fischer}, D. 2010, \apj, 725, 875

\bibitem[{{Janson} {et~al.}(2012){Janson}, {Carson}, {Lafreni{\`e}re},
  {Spiegel}, {Bent}, \& {Wong}}]{janson12}
{Janson}, M., {Carson}, J.~C., {Lafreni{\`e}re}, D., {et~al.} 2012, \apj, 747,
  116

\bibitem[{{Kalas} {et~al.}(2010){Kalas}, {Fitzgerald}, {Papadopoulos},
  {Graham}, {Maness}, {Matthews}, {Clampin}, {Chiang}, \& {Basu}}]{kalas10}
{Kalas}, P., {Fitzgerald}, M.~P., {Papadopoulos}, P.~P., {et~al.} 2010, in
  American Astronomical Society Meeting Abstracts 215, Vol.~42, 377.07

\bibitem[{{Kalas} {et~al.}(2008){Kalas}, {Graham}, {Chiang}, {Fitzgerald},
  {Clampin}, {Kite}, {Stapelfeldt}, {Marois}, \& {Krist}}]{kalas08}
{Kalas}, P., {Graham}, J.~R., {Chiang}, E., {et~al.} 2008, Science, 322, 1345

\bibitem[{{Kalas} {et~al.}(2005){Kalas}, {Graham}, \& {Clampin}}]{kalas05}
{Kalas}, P., {Graham}, J.~R., \& {Clampin}, M. 2005, \nat, 435, 1067

\bibitem[{{Lagrange} {et~al.}(2000){Lagrange}, {Backman}, \&
  {Artymowicz}}]{lagrange00}
{Lagrange}, A.-M., {Backman}, D.~E., \& {Artymowicz}, P. 2000, Protostars and
  Planets IV, 639

\bibitem[{{Lagrange} {et~al.}(2010){Lagrange}, {Bonnefoy}, {Chauvin}, {Apai},
  {Ehrenreich}, {Boccaletti}, {Gratadour}, {Rouan}, {Mouillet}, {Lacour}, \&
  {Kasper}}]{lagrange10}
{Lagrange}, A.-M., {Bonnefoy}, M., {Chauvin}, G., {et~al.} 2010, Science, 329,
  57

\bibitem[{{Lagrange} {et~al.}(2012){Lagrange}, {De Bondt}, {Meunier},
  {Sterzik}, {Beust}, \& {Galland}}]{lagrange12}
{Lagrange}, A.-M., {De Bondt}, K., {Meunier}, N., {et~al.} 2012, \aap, 542, A18

\bibitem[{{Lagrange} {et~al.}(2009{\natexlab{a}}){Lagrange}, {Desort},
  {Galland}, {Udry}, \& {Mayor}}]{lagrange09b}
{Lagrange}, A.-M., {Desort}, M., {Galland}, F., {Udry}, S., \& {Mayor}, M.
  2009{\natexlab{a}}, \aap, 495, 335

\bibitem[{{Lagrange} {et~al.}(2009{\natexlab{b}}){Lagrange}, {Gratadour},
  {Chauvin}, {Fusco}, {Ehrenreich}, {Mouillet}, {Rousset}, {Rouan}, {Allard},
  {Gendron}, {Charton}, {Mugnier}, {Rabou}, {Montri}, \&
  {Lacombe}}]{lagrange09a}
{Lagrange}, A.-M., {Gratadour}, D., {Chauvin}, G., {et~al.} 2009{\natexlab{b}},
  \aap, 493, L21

\bibitem[{{Lanza} {et~al.}(2011){Lanza}, {Boisse}, {Bouchy}, {Bonomo}, \&
  {Moutou}}]{lanza11}
{Lanza}, A.~F., {Boisse}, I., {Bouchy}, F., {Bonomo}, A.~S., \& {Moutou}, C.
  2011, \aap, 533, A44

\bibitem[{{Lebreton} {et~al.}(2013){Lebreton}, { van Lieshout}, { Augereau}, {
  Absil}, { Mennesson}, { Kama}, { Dominik}, { Bonsor}, { Beust}, { Defrere},
  {Faramaz}, { Hinz}, { Kral}, { Lagrange}, { Liu}, \&
  {Thebault}}]{lebreton13b}
{Lebreton}, J., { van Lieshout}, R., { Augereau}, J., {et~al.} 2013, submitted
  to A\&A, 1, 1

\bibitem[{{Lebreton} {et~al.}(2012){Lebreton}, {Augereau}, {Thi}, {Roberge},
  {Donaldson}, {Schneider}, {Maddison}, {M{\'e}nard}, {Riviere-Marichalar},
  {Mathews}, {Kamp}, {Pinte}, {Dent}, {Barrado}, {Duch{\^e}ne}, {Gonzalez},
  {Grady}, {Meeus}, {Pantin}, {Williams}, \& {Woitke}}]{lebreton12}
{Lebreton}, J., {Augereau}, J.-C., {Thi}, W.-F., {et~al.} 2012, \aap, 539, A17

\bibitem[{{Lockwood} {et~al.}(2007){Lockwood}, {Skiff}, {Henry}, {Henry},
  {Radick}, {Baliunas}, {Donahue}, \& {Soon}}]{lockwood07}
{Lockwood}, G.~W., {Skiff}, B.~A., {Henry}, G.~W., {et~al.} 2007, \apjs, 171,
  260

\bibitem[{{Macintosh} {et~al.}(2008){Macintosh}, {Graham}, {Palmer}, {Doyon},
  {Dunn}, {Gavel}, {Larkin}, {Oppenheimer}, {Saddlemyer}, {Sivaramakrishnan},
  {Wallace}, {Bauman}, {Erickson}, {Marois}, {Poyneer}, \&
  {Soummer}}]{macintosh08}
{Macintosh}, B.~A., {Graham}, J.~R., {Palmer}, D.~W., {et~al.} 2008, in Society
  of Photo-Optical Instrumentation Engineers (SPIE) Conference Series, Vol.
  7015

\bibitem[{{Mamajek}(2012)}]{mama12}
{Mamajek}, E.~E. 2012, \apjl, 754, L20

\bibitem[{{Marois} {et~al.}(2008){Marois}, {Macintosh}, {Barman}, {Zuckerman},
  {Song}, {Patience}, {Lafreni{\`e}re}, \& {Doyon}}]{marois08}
{Marois}, C., {Macintosh}, B., {Barman}, T., {et~al.} 2008, Science, 322, 1348

\bibitem[{{Marois} {et~al.}(2010){Marois}, {Zuckerman}, {Konopacky},
  {Macintosh}, \& {Barman}}]{marois10}
{Marois}, C., {Zuckerman}, B., {Konopacky}, Q.~M., {Macintosh}, B., \&
  {Barman}, T. 2010, \nat, 468, 1080

\bibitem[{{Mayor} {et~al.}(2011){Mayor}, {Marmier}, {Lovis}, {Udry},
  {S{\'e}gransan}, {Pepe}, {Benz}, {Bertaux}, {Bouchy}, {Dumusque}, {Lo Curto},
  {Mordasini}, {Queloz}, \& {Santos}}]{mayor11}
{Mayor}, M., {Marmier}, M., {Lovis}, C., {et~al.} 2011, ArXiv e-prints

\bibitem[{{Mayor} {et~al.}(2003){Mayor}, {Pepe}, {Queloz}, {Bouchy},
  {Rupprecht}, {Lo Curto}, {Avila}, {Benz}, {Bertaux}, {Bonfils}, {Dall},
  {Dekker}, {Delabre}, {Eckert}, {Fleury}, {Gilliotte}, {Gojak}, {Guzman},
  {Kohler}, {Lizon}, {Longinotti}, {Lovis}, {Megevand}, {Pasquini}, {Reyes},
  {Sivan}, {Sosnowska}, {Soto}, {Udry}, {van Kesteren}, {Weber}, \&
  {Weilenmann}}]{mayor03}
{Mayor}, M., {Pepe}, F., {Queloz}, D., {et~al.} 2003, The Messenger, 114, 20

\bibitem[{{Mesa} {et~al.}(2011){Mesa}, {Gratton}, {Berton}, {Antichi},
  {Verinaud}, {Boccaletti}, {Kasper}, {Claudi}, {Desidera}, {Giro}, {Beuzit},
  {Dohlen}, {Feldt}, {Mouillet}, {Chauvin}, \& {Vigan}}]{mesa11}
{Mesa}, D., {Gratton}, R., {Berton}, A., {et~al.} 2011, \aap, 529, A131

\bibitem[{{Meunier} \& {Lagrange}(2013)}]{meunier13}
{Meunier}, N. \& {Lagrange}, A.-M. 2013, \aap, 551, A101

\bibitem[{{Meunier} {et~al.}(2012){Meunier}, {Lagrange}, \& {De
  Bondt}}]{meunier12}
{Meunier}, N., {Lagrange}, A.-M., \& {De Bondt}, K. 2012, \aap, 545, A87

\bibitem[{{Mo{\'o}r} {et~al.}(2006){Mo{\'o}r}, {{\'A}brah{\'a}m}, {Derekas},
  {Kiss}, {Kiss}, {Apai}, {Grady}, \& {Henning}}]{moor06}
{Mo{\'o}r}, A., {{\'A}brah{\'a}m}, P., {Derekas}, A., {et~al.} 2006, \apj, 644,
  525

\bibitem[{{Patience} {et~al.}(2011){Patience}, {Bulger}, {King}, {Ayliffe},
  {Bate}, {Song}, {Pinte}, {Koda}, {Dowell}, \& {Kov{\'a}cs}}]{patience11}
{Patience}, J., {Bulger}, J., {King}, R.~R., {et~al.} 2011, \aap, 531, L17

\bibitem[{{Paulson} {et~al.}(2004){Paulson}, {Cochran}, \&
  {Hatzes}}]{paulson04}
{Paulson}, D.~B., {Cochran}, W.~D., \& {Hatzes}, A.~P. 2004, \aj, 127, 3579

\bibitem[{{Paulson} \& {Yelda}(2006)}]{paulson06}
{Paulson}, D.~B. \& {Yelda}, S. 2006, \pasp, 118, 706

\bibitem[{{Rameau} {et~al.}(2013){Rameau}, {Chauvin}, {Lagrange}, {Klahr},
  {Bonnefoy}, {Mordasini}, {Bonavita}, {Desidera}, {Dumas}, \&
  {Girard}}]{rameau13}
{Rameau}, J., {Chauvin}, G., {Lagrange}, A.-M., {et~al.} 2013, ArXiv e-prints

\bibitem[{{Rhee} {et~al.}(2007){Rhee}, {Song}, {Zuckerman}, \&
  {McElwain}}]{rhee07}
{Rhee}, J.~H., {Song}, I., {Zuckerman}, B., \& {McElwain}, M. 2007, \apj, 660,
  1556

\bibitem[{{Rodriguez} \& {Zuckerman}(2012)}]{rodriguez12}
{Rodriguez}, D.~R. \& {Zuckerman}, B. 2012, \apj, 745, 147

\bibitem[{{Santos} {et~al.}(2010){Santos}, {Gomes da Silva}, {Lovis}, \&
  {Melo}}]{santos10}
{Santos}, N.~C., {Gomes da Silva}, J., {Lovis}, C., \& {Melo}, C. 2010, \aap,
  511, A54

\bibitem[{{Schneider} {et~al.}(2006){Schneider}, {Silverstone}, {Hines},
  {Augereau}, {Pinte}, {M{\'e}nard}, {Krist}, {Clampin}, {Grady}, {Golimowski},
  {Ardila}, {Henning}, {Wolf}, \& {Rodmann}}]{schneider06}
{Schneider}, G., {Silverstone}, M.~D., {Hines}, D.~C., {et~al.} 2006, \apj,
  650, 414

\bibitem[{{Schr{\"o}der} {et~al.}(2009){Schr{\"o}der}, {Reiners}, \&
  {Schmitt}}]{schroeder09}
{Schr{\"o}der}, C., {Reiners}, A., \& {Schmitt}, J.~H.~M.~M. 2009, \aap, 493,
  1099

\bibitem[{{Setiawan} {et~al.}(2008){Setiawan}, {Henning}, {Launhardt},
  {M{\"u}ller}, {Weise}, \& {K{\"u}rster}}]{setiawan08}
{Setiawan}, J., {Henning}, T., {Launhardt}, R., {et~al.} 2008, \nat, 451, 38

\bibitem[{{Setiawan} {et~al.}(2007){Setiawan}, {Weise}, {Henning}, {Launhardt},
  {M{\"u}ller}, \& {Rodmann}}]{setiawan07}
{Setiawan}, J., {Weise}, P., {Henning}, T., {et~al.} 2007, \apjl, 660, L145

\bibitem[{{Sozzetti}(2011)}]{sozzetti11}
{Sozzetti}, A. 2011, in EAS Publications Series, Vol.~45, EAS Publications
  Series, 273--278

\bibitem[{{Torres} {et~al.}(2008){Torres}, {Quast}, {Melo}, \&
  {Sterzik}}]{torres08}
{Torres}, C.~A.~O., {Quast}, G.~R., {Melo}, C.~H.~F., \& {Sterzik}, M.~F. 2008,
  {Young Nearby Loose Associations}, ed. B.~{Reipurth}, 757

\bibitem[{{Vican}(2012)}]{vican12}
{Vican}, L. 2012, \aj, 143, 135

\bibitem[{{Weise} {et~al.}(2010){Weise}, {Launhardt}, {Setiawan}, \&
  {Henning}}]{weise10}
{Weise}, P., {Launhardt}, R., {Setiawan}, J., \& {Henning}, T. 2010, \aap, 517,
  A88

\bibitem[{{Wright} {et~al.}(2004){Wright}, {Marcy}, {Butler}, \&
  {Vogt}}]{wright04}
{Wright}, J.~T., {Marcy}, G.~W., {Butler}, R.~P., \& {Vogt}, S.~S. 2004, \apjs,
  152, 261

\bibitem[{{Zuckerman} \& {Song}(2004)}]{zuckerman04}
{Zuckerman}, B. \& {Song}, I. 2004, \araa, 42, 685

\bibitem[{{Zuckerman} {et~al.}(2001){Zuckerman}, {Song}, {Bessell}, \&
  {Webb}}]{zuckerman01}
{Zuckerman}, B., {Song}, I., {Bessell}, M.~S., \& {Webb}, R.~A. 2001, \apjl,
  562, L87

\end{thebibliography}

\clearpage

\begin{landscape} 
\centering
\begin{table}[htp!]
\caption{Target list and stellar parameters.}
\label{targets}  
\begin{tabular*}{1.2\textwidth}{lllllllllllllll}
\hline
HIP      &         HD      &    Name   &    RA        & Dec         &    V    &    B-V  &   SpT   &   d     &  \vsini  & log(R'$_{HK}$) & Mass        & Age      &   MG           & Disk   \\
         &                 &           &   (2000)     &  (2000)     & (mag)   &         &         &   (pc)  &  (km/s)      &               & (M$_{\odot}$) & (Myr)    &                &        \\
\hline
  1113  &          987   &           & 00 13 53     &  -74 41 18   &    8.78 &    0.74 & G8V    &    44.4  &  7.3       &           & 0.98       & 30        & TucHor     & yes\\
   26373 &        37572    &UY Pic     & 05 36 56.85  &  -47 57 52  &    7.95 &    0.85 & K0V    &    25.1  &              & -4.20/-4.23   & 0.93       & 70        & AB Dor         & no     \\
   27321 &         39060   &$\beta$Pic & 05 47 17.1   &  -51  O3 59 &    3.85 &    0.17 & A3V    &    19.4  &              &               & 1.64       & 12        & $\beta$ Pic     & yes    \\
         &         42270   &           & 05 53 29     &  -81 56 53  &    9.0  &    0.77 & K0V    &          &              &               &  1.0       & 30        & Car            & no     \\
   30314 &         45270   &           & 06 22 30.94  &  -60 13 70  &    6.53 &    0.61 & G1V    &    23.8  &              &               & 1.11       & 70        & AB Dor         & no     \\
   36948 &         61005   &           & 07 35 47.46  &  -32 12 14  &    8.22 &    0.74 & G8V    &      35  &  9.9/8.2     &    -4.32      &            & 40$^a$    & Argus          & yes    \\
   41307 &         71155   &     30Mon & 08 25 39.63  &  -03 54 23  &    3.91 &    0.01 & A0V    &    37.5  &              &               & 2.23       & 170$^b$   & Field          & yes \\
   51386 &         90905   &           & 10 29 42.23  &   01 29 28  &    6.88 &    0.56 & F5     &    31.6  & 10.0         &  -4.43        & 1.16       & 400$^a$   & Field          &     no    \\
    57524&        102458   &  TWA19A   & 11 47 24.545 &  -49 53 03  &    9.07 &    0.63 & G4V    &   91.6   &              &               &            & 8         & TWA            &  no     \\
   59315 &        105690   &           & 12 10 06.47  &  -49 10 50  &    8.16 &    0.71 & G5V    &     37.8 & 8.0/8.5      &  -4.27        & 1.02       & 230$^a$   & Field          &   no   \\
   61468 &        109536   &           & 12 35 45.53  &  -41  1 19  &    5.12 &    0.22 & A7III  &     34.6 &              &               & 1.63       & 810$^b$   &  Field         &  no\\
   74405 &        133813   &           & 15 12 23     &  -75 15 16  &     9.42&    0.81 & G9     &     50.3 &              &               &            & 40        & Argus          & no\\
         &   141943        &    NZ Lup & 15 53 27.3   &  -42 16 10  &    7.92 &    0.65 & G2     &     43.8 & 36.3/40.0    &   -4.28/-4.36 & 1.09       & 24.4$^a$  & Field          &  no      \\
   79881 &        146624   &           & 16 18 17.9   &  -28 36 51  &    4.8  &    0.01 & A0V    &     41.3 &              &               & 1.9        & 12        & $\beta$ Pic     &   no \\
   92024 &        172555   &   HR 7012 & 18 45 26.9   &  -64 52 16  &    4.78 &    0.2  & A7V    &     28.5 &              &               & 1.61       & 12        &  $\beta$ Pic    & yes    \\
   92680 &        174429   &    PZ Tel & 18 53 5.9    &  -50 10 50  &    8.29 &    0.77 & G9IV   &     51.5 &              &               & 1.1        & 12        & $\beta$ Pic     &  no \\
  93815	 &	177171	   & rho Tel   & 19 06 19.96  & -52 20 27.27&    5.18 & 0.50	& F7V	 &     52.6 &	       	   &		   &		& 30        & TucHor         & no\\
   95149 &        181327   &           & 19 22 58.94  &  -54 32 17  &    7.04 &    0.46 & F6V    &     50.6 &  19.5        &               &   1.36     & 12        & $\beta$ Pic     & yes  \\
   96334 &        183414   &           & 19 35 9.72   &  -69 58 32  &    7.89 &    0.65 & G3V    &     35.4 &  9.8         &  -4.23        & 1.04       & 192$^a$   & Field          & yes           \\
   98495 &        188228   &    eps Pav& 20 0 35.55   &  -72 54 38  &    3.97 &   -0.03 & A0V    &     32.2 &              &               & 2.03       & 40        & Argus          &  no\\
   102626&        197890   &    BO Mic & 20 47 44.97  &  -36 35 41  &    9.4  &    0.94 & K3Ve   &     52.2     & 140.0        &               &            & 30        & TucHor         & no\\
  107947 &        207575   &           & 21 52  9.7   &  -62  3  80 &    7.22 &    0.51 & F6V    &36.5/45.3 &              &               & 1.24       & 30        & TucHor         & no \\
  113368 &        216956   & Fomalhaut & 22 57 39.05  &  -29 37 20  &    1.17 &    0.14 & A3V    &      7.7 &              &               & 1.89       & 440$^c$   & Field          & yes         \\
  113579 &        217343   &           & 23 0 19.29   &  -26  9 13  &    7.47 &    0.65 & G5V    &     30.8 & 12.2/12.4    &  -4.29        & 1.05       & 70        & AB Dor         & no   \\
  114189 &        218396   &   HR 8799 & 23 7 28.72   &   21  8  30 &    5.97 &    0.26 & A5V    &     39.4 &              &               & 1.46       & 30        & Col            & yes    \\
  11808  &        224228   &           & 23 56 10.67  &  -39  3  80 &    8.22 &    0.97 & K3V    &     22.0 &     8.7      &               & 0.86       & 70        & AB Dor         & no   \\
\hline
 \end{tabular*}
\tablefoot{SpT refers to spectral type, MG stands for Moving Group while
  Disk indicates the presence of a dusty disk. The \vsini are from
  \cite{weise10}, except for HD90905 \cite[][]{wright04}. The
  log(R'$_{HK}$) are from \cite{wright04} except for HD61005
  \cite[][]{schroeder09}. Ages are from Zuckerman \& Song (2004) and Torres et (2008) for the moving
  groups members of AB Dor, $\beta$ Pic, Tucana-Horologium (TucHor),
  Columba (Col), Carina (Car) and Argus, except when labelled $^a$
  \cite[][]{desidera11}, $^b$ \cite[][]{vican12} and $^c$\cite[][]{mama12}. Young field stars
  are flagged as ''Field''.}
\end{table}
\end{landscape}


\begin{landscape}
\centering
      \begin{table}[htp!]
	\caption{Observation results and typical detection limits.}
	\label{obs}
\begin{tabular*}{1.25\textwidth}{llllllllllllll}
	\hline
HD	&	N	&	Time span	&	RV rms	&	RV unc	&	RV amp 	&	BVS rms	&	BVS unc	&	BVS  ampl 	&	Bisector	&	\vsini &	Ca line	&	Variab. type	&	Det. limits 	\\
\hline
	&		&	(days)	&	(m/s)	&	(m/s)	&	(m/s)	&	(m/s)	&	(m/s)	&	(m/s)	&		&	(km/s)	&		&		&	(\Mjup)	\\
987	&	11	&	39.0	&	90.9	&	2.5	&	326.5	&	95.4	&	6.3	&	338.1	&	G	&	 7.3	&	strong (-4.06)	&	A	&		\\
37572	&	15	&	222.4	&	65.1	&	1.5	&	221.2	&	59.4	&	4.1	&	221.2	&	G	&	9.6	&	strong (-4.27)	&	A	&		\\
39060	&	1049	&	2,656.8	&	275.2	&	39.1	&	1680.0	&	783.4	&	97.7	&	7387.0	&	A	&		&	absorption	&	HFV/P	&	1.5/2.1/4/12.4/16	\\
42270	&	2	&	0.1	&	82.7	&	11.8	&	154.6	&	96.8	&	30.5	&	193.7	&	G	&	31.2	&	strong ($\geq$-4.0)	&	NEP	&		\\
45270	&	13	&	189.3	&	33.9	&	1.7	&	126.2	&	73.9	&	4.4	&	288.5	&	G	&	17.7	&	small	&	A	&		\\
61005	&	12	&	4.0	&	40.1	&	2.6	&	104.7	&	32.9	&	8.3	&	96.7	&	G	&	9.7	&	medium (-4.20)	&	A	&		\\
71155	&	76	&	1,168.9	&	451.8	&	230.7	&	2454.3	&		&		&		&	B 	&		&	no	&	HFV/P	&	8.8/18/33/46/49	\\
90905	&	13	&	45.1	&	48.4	&	1.3	&	126.7	&	65.3	&	3.5	&	182.7	&	G	&	9.5	&	small (-4.29)	&	A	&		\\
102458	&	16	&	17.2	&	331.7	&	9.8	&	888.9	&	491.7	&	24.5	&	1274.0	&	G	&	28.1	&	medium	&	A	&		\\
105690	&	104	&	823.0	&	58.9	&	2.5	&	232.0	&	50.1	&	6.1	&	213.7	&	G	&	9.6	&	medium/small (-4.25)	&	A	&	0.4/0.5/1.2/2/-	\\
109536	&	46	&	93.0	&	214.6	&	29.2	&	791.3	&	762.7	&	73.0	&	2987.7	&	G	&	88.4	&	no	&	HFV/P	&	3.6/1.2/10/-/-	\\
133813	&	2	&	0.0	&	2.2	&	5.2	&	4.0	&	20.6	&	13.1	&	41.2	&	G	&	13.3	&	strong (-4.0)	&	NEP	&		\\
141943	&	38	&	149.9	&	235.5	&	8.0	&	900.5	&	299.9	&	20.0	&	1113.7	&	G	&	34.9	&	strong ($\geq$-4.0)	&	A (-1.1)	&	1.5/3.3/5.5/-/-	\\
146624	&	25	&	1,878.9	&	20.4	&	16.2	&	95.1	&		&		&		&	G	&		&	no	&	HFV/P	&		\\
172555	&	49	&	1,887.9	&	264.1	&	58.7	&	1299.5	&		&		&		&	N/A	&		&	absorption	&	HFV/P 	&	3.4/4.4/12.5/15.7/34	\\
174429	&	22	&	1,568.9	&	407.3	&	27.5	&	1361.3	&	112.9	&	68.9	&	1388.8	&	A	&	79.6	&	strong ($\geq$-4.0)	&	A	&		\\
177171	&	28	&	143.8	&	1435.8	&	9.9	&	5682.0	&		&		&		&	B 	&		&		&	SB	&		\\
181327	&	40	&	587.7	&	15.1	&	3.2	&	55.5	&	27.4	&	8.1	&	129.1	&	G	&	17.9	&	no 	&	A (-1.1)	&	0.2/0.3/0.5/1.5/-	\\
183414	&	41	&	188.9	&	66.7	&	2.2	&	204.4	&	69.4	&	5.3	&	226.2	&	G	&	11.1	&	small (-4.23)	&	A (-1)	&	0.7/0.6/1.7/-/-	\\
188228	&	331	&	1,983.8	&	127.3	&	88.9	&	719.4	&       	&	       &		&	N/A	&		&	absorption	&	HFV/P 	&1.5/3.2/7.0/20/30			\\
197890	&	2	&	0.009	&	161.9	&	322.3	&	253.9	&		&		&		&	NEP	&		&		&	NEP	&		\\
207575	&	27	&	187.9	&	40.6	&	7.4	&	165.2	&	89.1	&	17.9	&	384.4	&	G	&	32.8	&	small  (-4.29)	&	HFV/P	&		\\
216956	&	284	&	1,853.0	&	56.1	&	15.9	&	618.4	&	363.6	&	39.7	&	2821.4	&	A	&		&	no	&	HFV/P 	&	1/0.8/2.1/3.3/1.9	\\
217343	&	16	&	144.8	&	104.5	&	2.5	&	262.4	&	97.2	&	6.2	&	276.5	&	G	&	13.3	&	medium/small (-4.18)	&	A	&		\\
218396	&	10	&	1,302.4	&	490.3	&	32.4	&	2057.4	&	1802.0	&	96.3	&	8276.0	&	A	&		&	no	&	NEP	&		\\
224228	&	14	&	39.0	&	2.5	&	0.9	&	7.1	&	3.0	&	2.5	&	9.6	&	G	&	5.7	&	strong (-4.30)	&	constant	&		\\
\hline
   \end{tabular*}
	\tablefoot{
N is the number of spectra available for each target star.
``RV rms'' (respectively ``BVS rms'') is the rms of the measured radial velocities (respectively bisector velocity-spans).
``RV amp'' (respectively ``BVS amp.'') is the amplitude of the measured radial velocities (respectively bisector velocity-spans).
``RV unc'' (respectively ``BVS unc'') is the average uncertainty associated to the RV (respectively bisector velocity-spans) data. 
The column labelled "bisector" provides a flag of the BVS quality: good (G), acceptable (A), bad (B), not enough point (NEP). 
The \vsini have been computed from our data (see section 3.2). 
The column labelled "Ca line" provides a classification of the Ca emission with respect to the stellar line depth as follows: strong, medium, small, no, absorption Ca line. The mean log(R'$_{HK})$ is indicated between parenthesis whenever possible.  
The variability classification is as follows: active (A), pulsating (P), high frequency RV variations (HFV), spectroscopic binary (SB), candidate companion (CC), not enough points (NEP).
The slope of the (RV, BVS) is indicated between parenthesis for the most active stars.
The detection limits are computed for periods of  3, 10, 100, 500 and 1000 days (whenever possible) using the LPA method on 
non-averaged data.}
\end{table}

\end{landscape}

\Online
\begin{appendix}
\section{Observed RV, BVS and measured detection limits}
\begin{figure*}[t!]
\centering
\includegraphics[angle=90,width=1.\hsize]{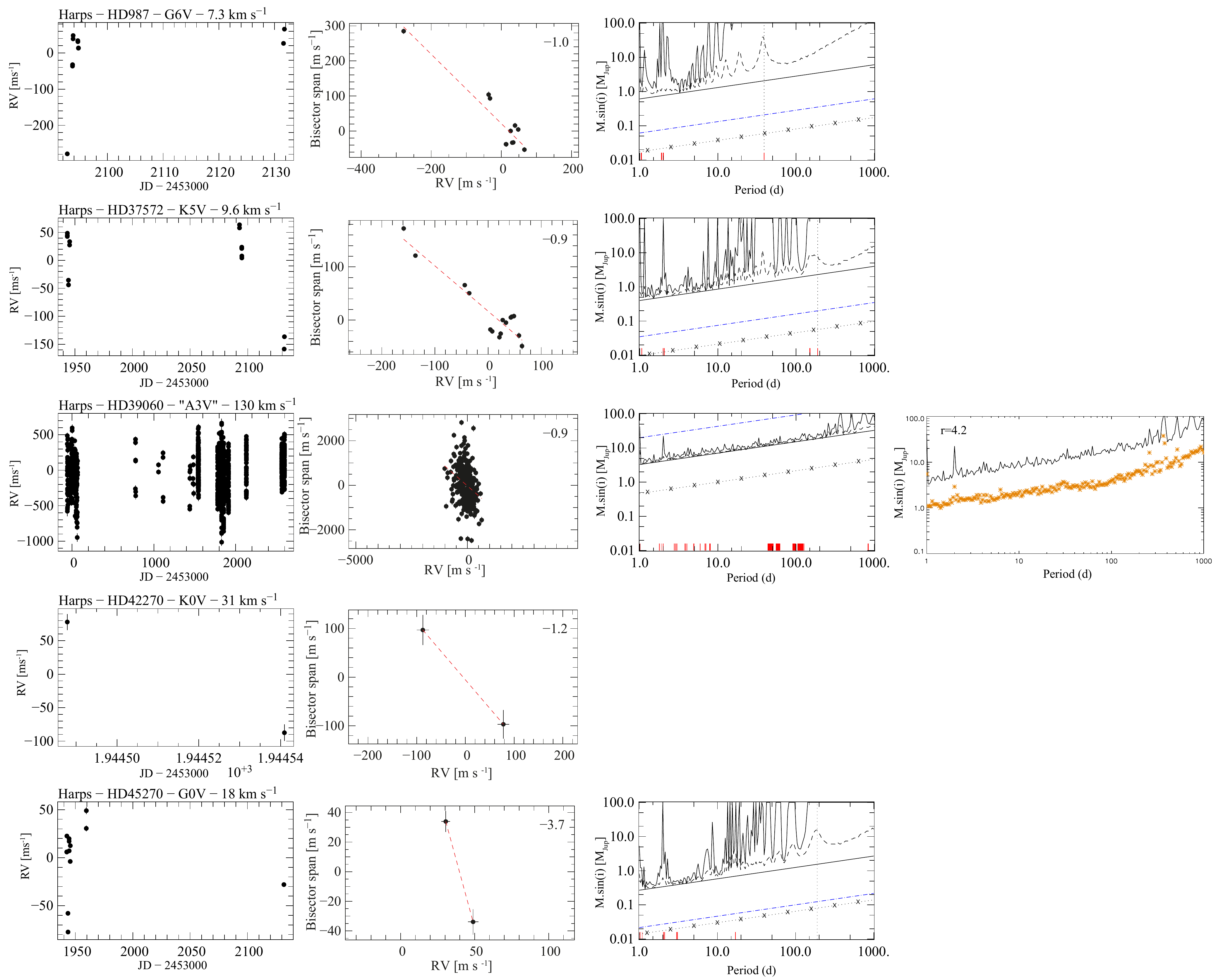}
\caption{ From Left to Right: 1) RV time series. 2) (RV, BVS) diagramme.  The slope of the (RV, BVS) correlation is indicated.  Note that in a few cases, the value is obviously impacted by a single deviating data point. 3) rms-based detection limits with a 99.8$\%$ probability (solid black curve) and a 68.2$\%$ probability (dotted curve), achievable limits with a perfect sampling (black solid straight line) and achievable limits if the signal was limited only by photon and instrumental noises and with a perfect temporal sampling (dotted straight lines). The vertical lines in the (period, detection limit) diagramme indicate the time-span for each star. 4) detection limits obtained with the LPA methods. "r" indicates the factor of improvement between the LPA and rms-based detection limits.}
\label{rv_limdet_1}
 \end{figure*}


\begin{figure*}[t!]
\centering
\includegraphics[angle=90,width=\hsize]{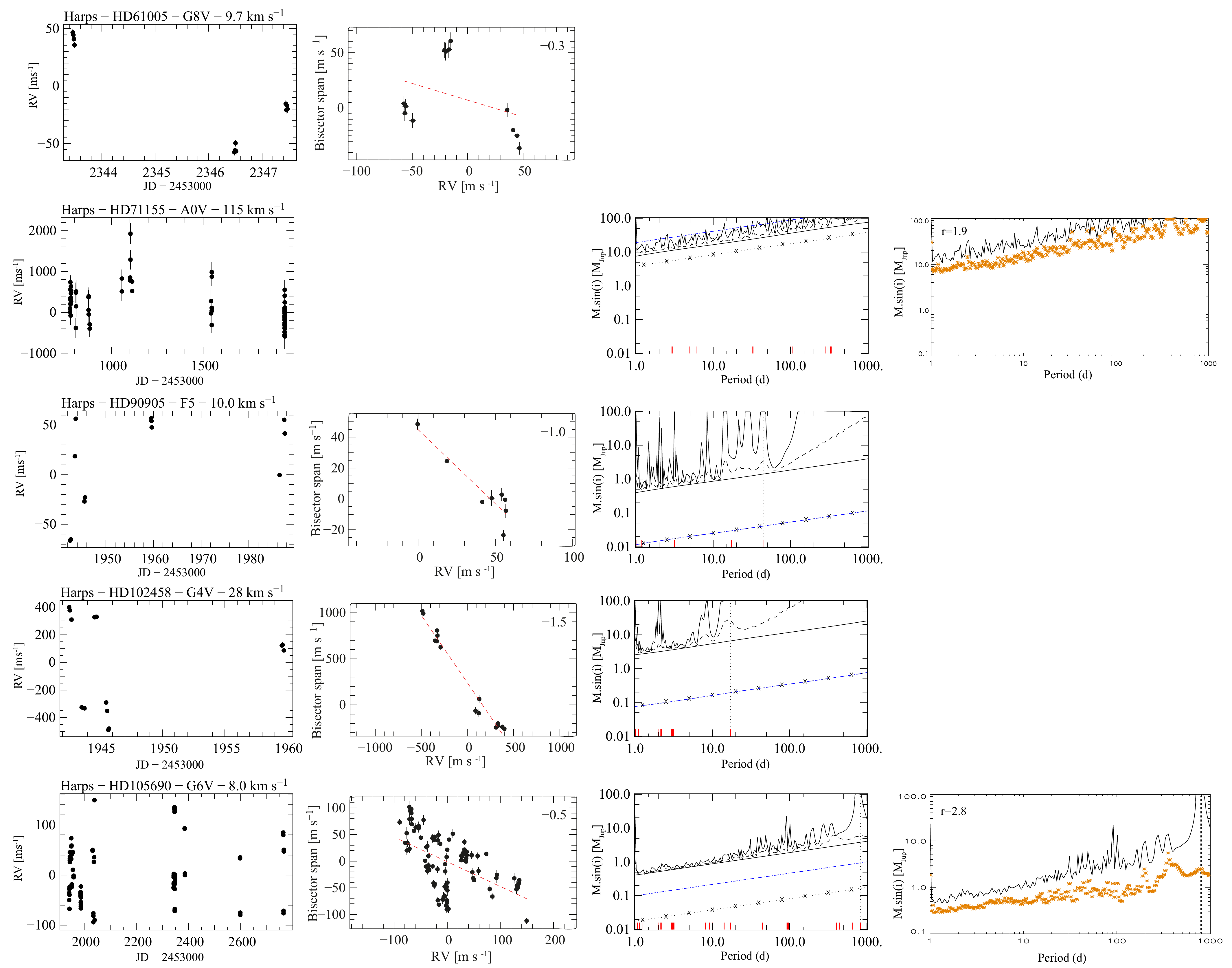}
\caption{Cont'd.}
\label{rv_limdet_2}
 \end{figure*}

\begin{figure*}[t!]
\centering
\includegraphics[angle=90,width=\hsize]{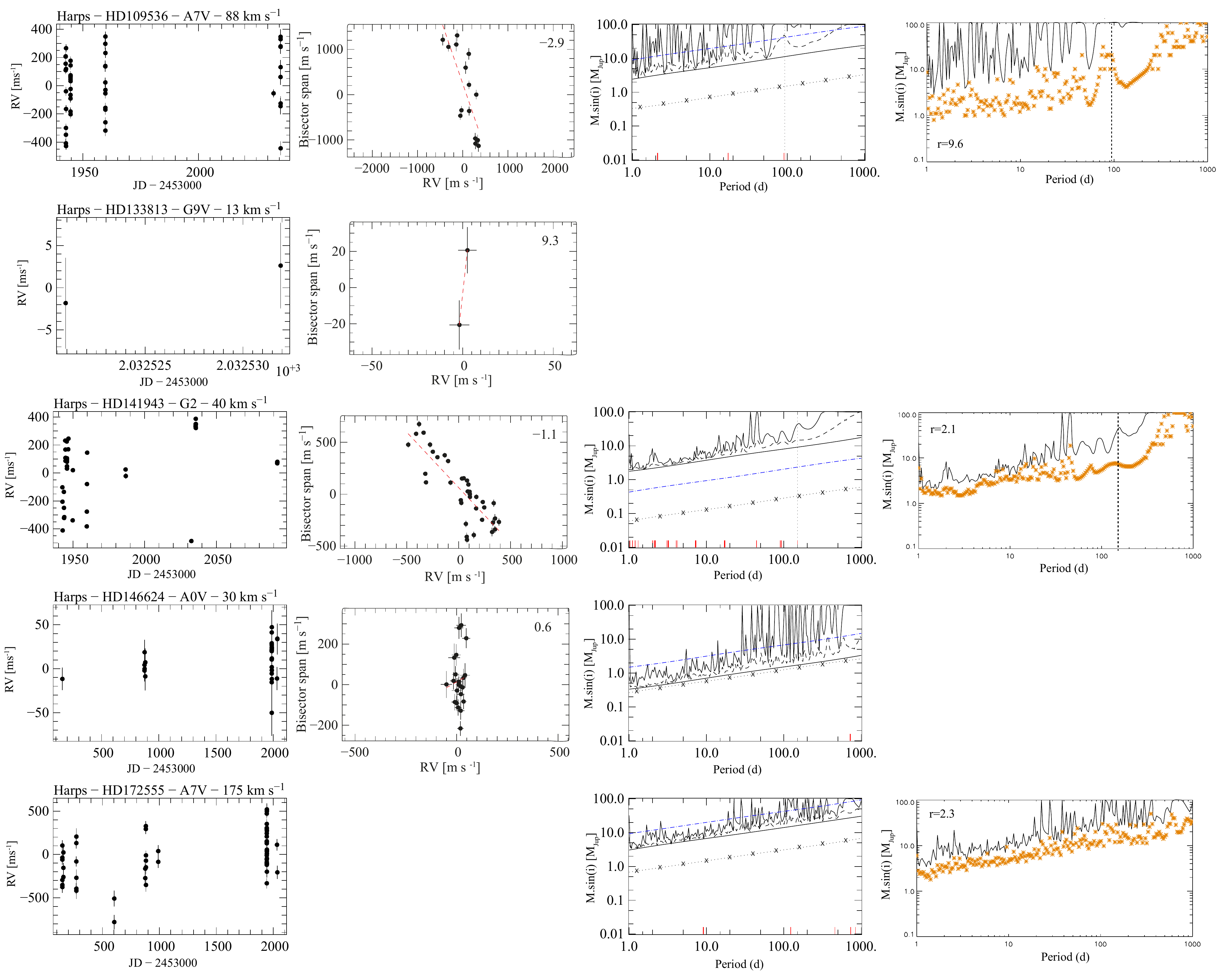}\\
\caption{Cont'd.}
\label{rv_limdet_3}
 \end{figure*}

\begin{figure*}[t!]
\centering
\includegraphics[angle=90,width=\hsize]{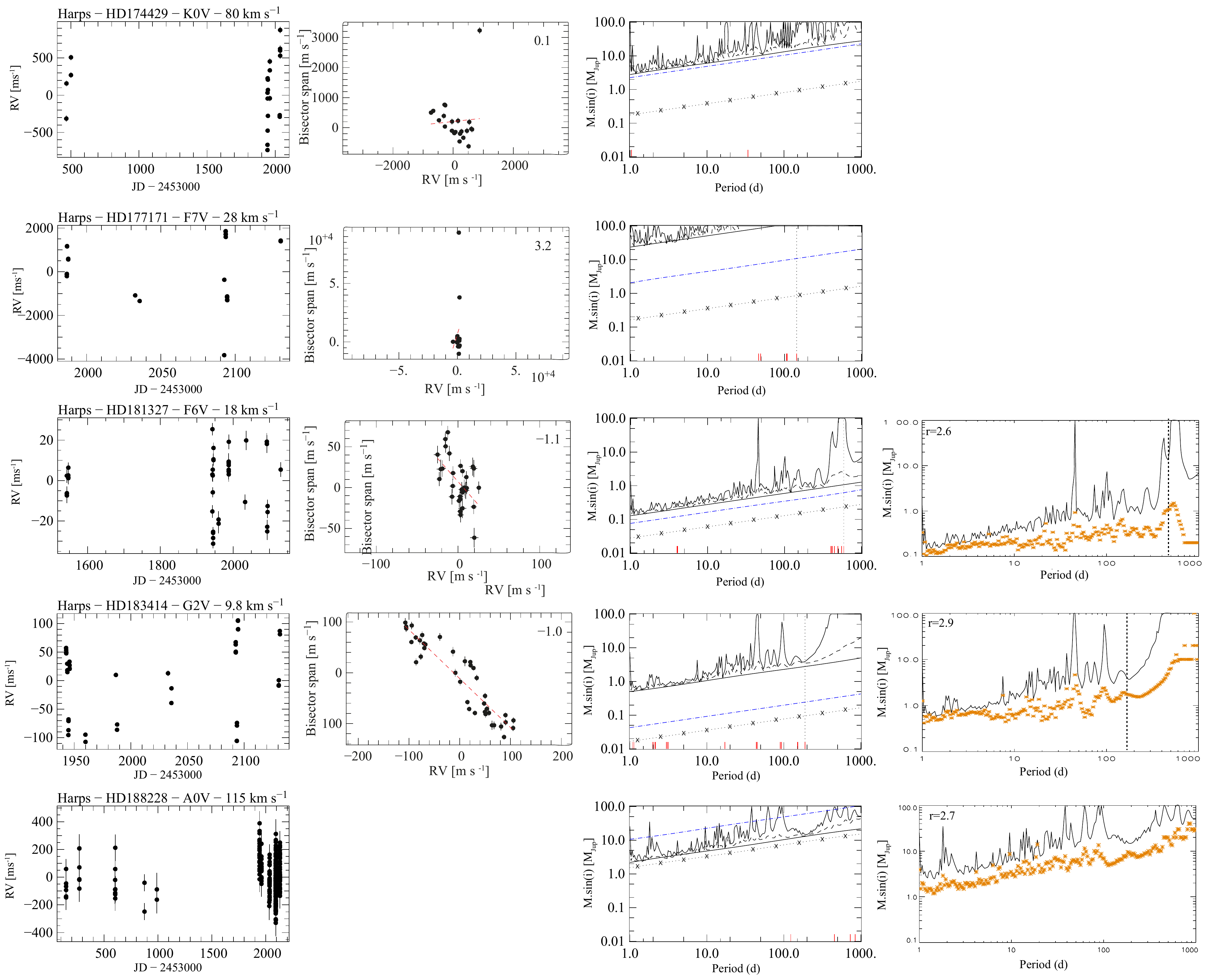}\\
\caption{Cont'd.}
\label{rv_limdet_4}
 \end{figure*}

\begin{figure*}[t!]
\centering
\includegraphics[angle=90,width=\hsize]{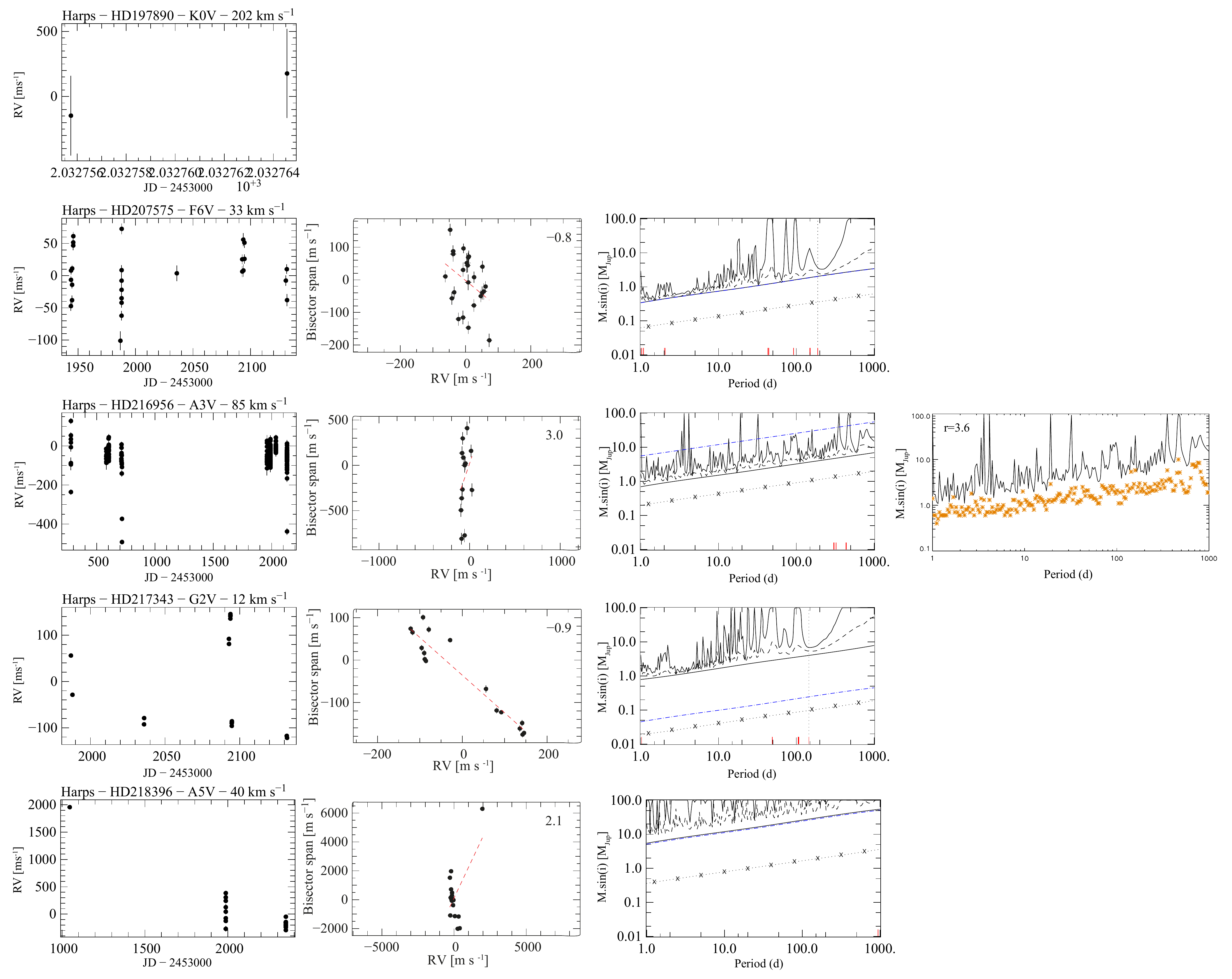}
\caption{Cont'd.}
\label{rv_limdet_5}
 \end{figure*}

\begin{figure*}[t!]
\centering
\includegraphics[angle=90,width=.3\hsize]{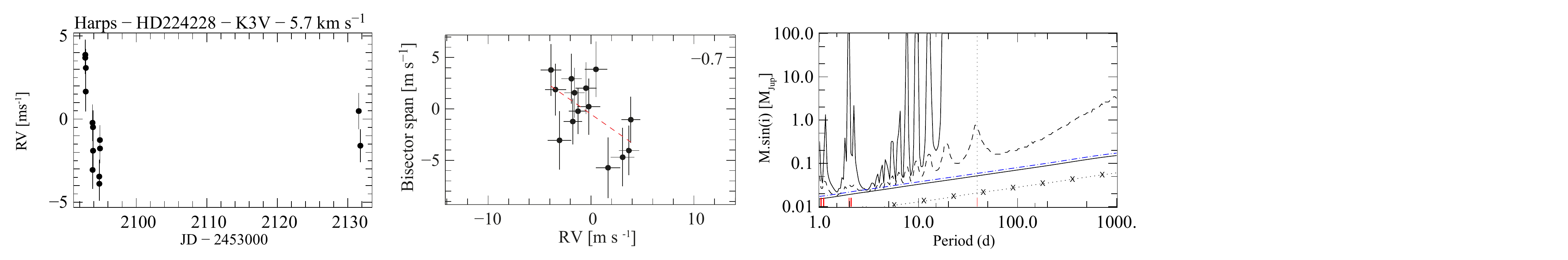}
\caption{Cont'd.}
\label{rv_limdet_6}
 \end{figure*}


\section{Simulations results}
\begin{figure*}[htp!]
    \centering
\includegraphics[angle=0,width=1.\hsize]{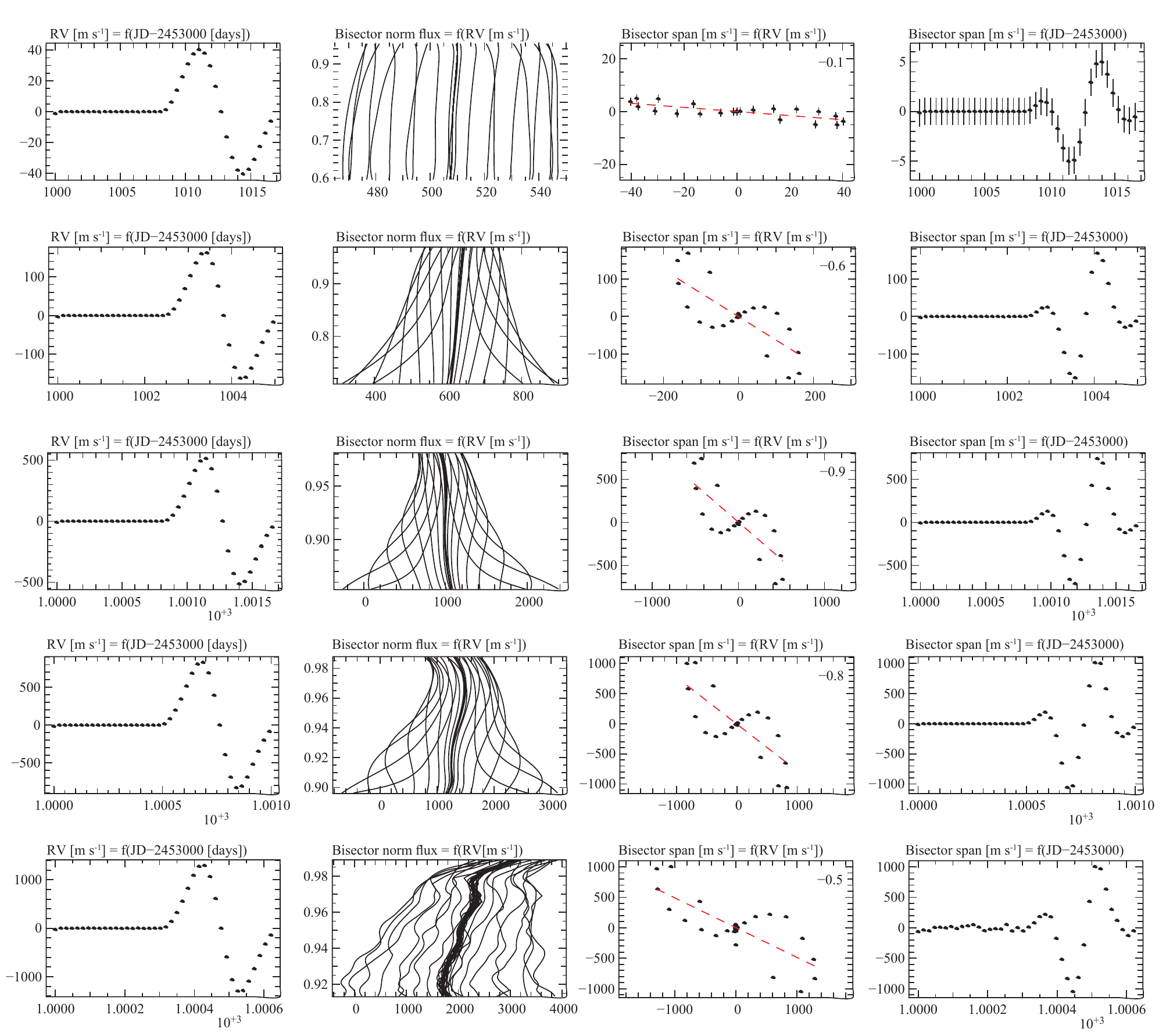}
    \caption{ {\it First column}: RV versus time of a simulated equatorial spot covering 1.5\% of the visible surface and a temperature contrast of 1200~K, for a seen edge-on sun-like star and for five different values of  \vsini from the upper panels to the lower panels: 3 km/s, 10 km/s, 30 km/s, 50 km/s, 80 km/s. {\it Second column}: same for the bissectors.  {\it Third column}: BVS versus RV. {\it Fourth column}: BVS versus time.}
    \label{eightshapes1}
  \end{figure*}

\begin{figure*}[htp!]
    \centering
\includegraphics[angle=0,width=1.\hsize]{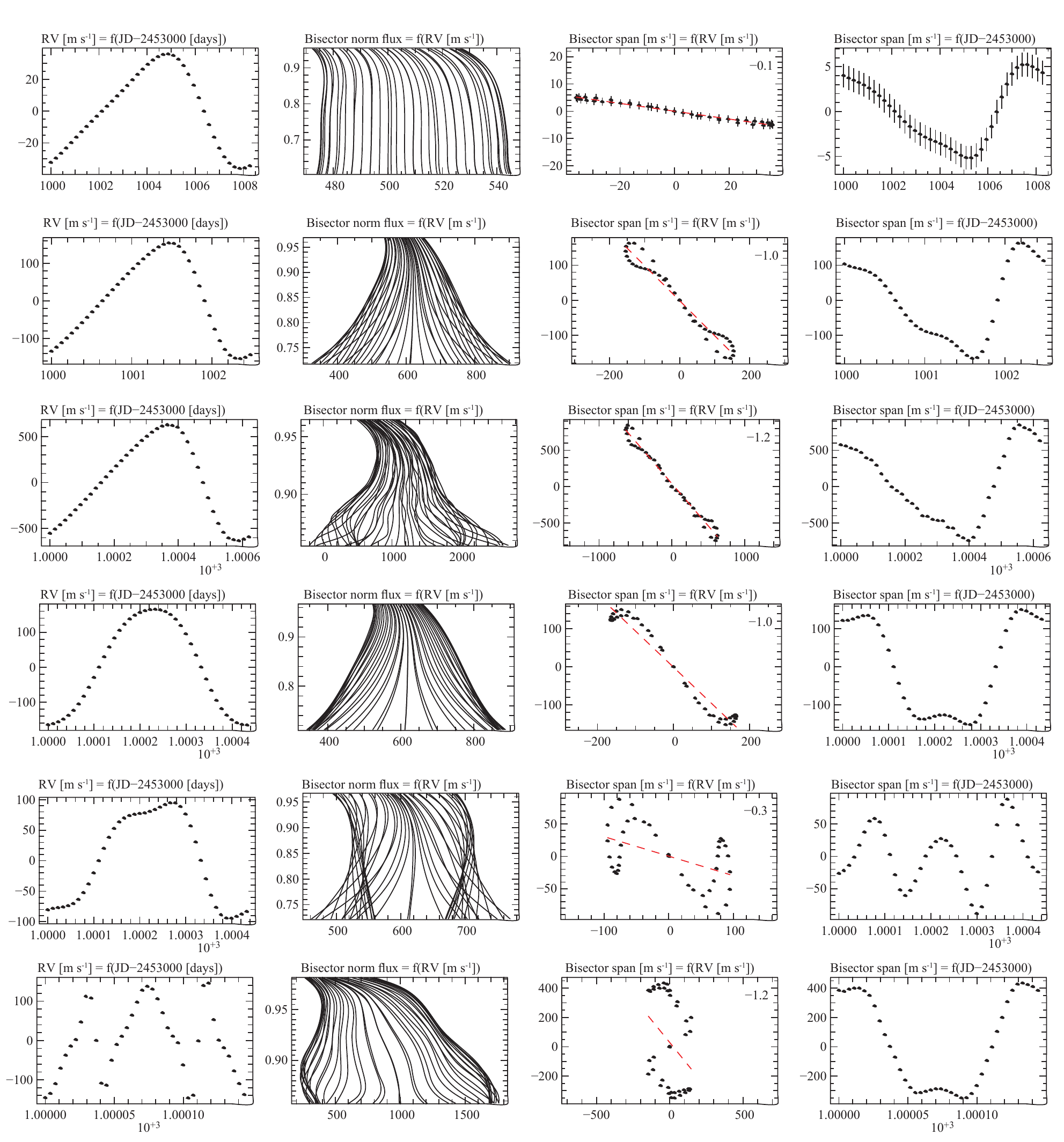}
    \caption{Same as Fig.~10 for an inclination of 30 degrees (three first lines) and 5 degrees (three last lines), a colatitude (from the pole) of 30 degrees (except 5th line with a colatitude of 60 degrees) and \vsini (from upper panels to lower panels) of 3, 10, 40, 10, 10 and 10 km/s.  }
    \label{eightshapes2}
  \end{figure*}

\end{appendix}

\vfill\eject

\end{document}